\documentclass[11pt,a4paper]{article}
\pdfoutput=1

\usepackage{cite}
\usepackage{physics}
\usepackage[english]{babel}
\usepackage[utf8]{inputenc}
\usepackage{amsmath}
\usepackage{graphicx}
\usepackage{tikz-feynman}
\usepackage[nottoc]{tocbibind}
\usepackage{amssymb}

\usepackage[colorlinks=true, allcolors=blue]{hyperref}
\usepackage{float}
\usepackage{multirow}
\usepackage{makeidx}

\numberwithin{figure}{section}
\numberwithin{equation}{section}

\newcommand{\be}{\begin{equation}}
\newcommand{\ee}{\end{equation}}
\newcommand{\bea}{\begin{eqnarray}}
\newcommand{\eea}{\end{eqnarray}}
\def\beal#1\eeal{\begin{align}#1\end{align}}   
\def\besp#1\eesp{\begin{multline}#1\end{multline}} 

\usepackage[normalem]{ulem}

\newcommand{\ignore}[1]{#1} 

\newcommand\ie{\textit{i.e.}\ }
\newcommand\eg{\textit{e.g.}\ }
\newcommand\cf{\textit{cf.}\ }

\newcommand{\aka}{{a.k.a.}\ }
\newcommand{\etc}{{\it etc.}\ }
\newcommand{\viz}{{\it viz.}\ }

\newcommand{\nn}{\nonumber}

\newcommand{\ph}{\varphi}
\newcommand{\vp}{\varphi}
\newcommand{\half}{\tfrac{1}{2}}

\newcommand{\eps}{\varepsilon}
\newcommand{\K}{\mathcal{K}}
\newcommand{\cO}{\mathcal{O}}
\newcommand{\R}{R^{(1)}}
\newcommand{\bR}{\bar{R}^{(1)}}

\begin{document}

\begin{titlepage}

\begin{center}
{\huge \bf Off-shell divergences in quantum gravity}
\end{center}
\vskip1cm


\begin{center}
{\bf Vlad-Mihai Mandric, Tim R. Morris and Dalius Stulga}
\end{center}

\begin{center}
{\it STAG Research Centre \& Department of Physics and Astronomy,\\  University of Southampton,
Highfield, Southampton, SO17 1BJ, U.K.}\\
\vspace*{0.3cm}
{\tt  V.M.Mandric@soton.ac.uk, T.R.Morris@soton.ac.uk, D.Stulga@soton.ac.uk}
\end{center}



\begin{abstract}
We investigate off-shell perturbative renormalisation of pure quantum gravity for both background metric and quantum fluctuations. We show that at each new loop order, the divergences that do not vanish on-shell are constructed from only the total metric, whilst those that vanish on-shell are renormalised by canonical transformations involving the quantum fields. Purely background metric divergences do not separately appear, and the background metric does not get renormalised. We highlight that renormalisation group identities play a crucial r\^ole ensuring consistency in the renormalisation of BRST transformations  beyond one loop order. We verify these assertions by computing leading off-shell divergences to two loops, exploiting off-shell BRST invariance and the renormalisation group equations. Although some divergences can be absorbed by field redefinitions, we  explain why this does not lead to finite beta-functions for the corresponding field.
\end{abstract}



\end{titlepage}
\tableofcontents
\newpage

\section{Introduction}
\label{sec:Introduction}

This paper is about four dimensional perturbative quantum gravity, constructed by quantising the Einstein-Hilbert action.
As is well known, this quantum field theory is not perturbatively renormalisable \cite{tHooft:1974toh,Goroff:1985sz,Goroff:1985th,vandeVen:1991gw}. At each new loop order $\ell$, counterterms have to be added to the bare action  to cancel  ultraviolet (UV) divergences, and associated with these counterterms are new operators and renormalised couplings that did not exist in the bare action at lower loop order. Nevertheless perturbative quantum gravity can be consistently treated as an effective theory in this way \cite{Gomis:1995jp}, see also \cite{Donoghue:1994dn,Burgess:2003jk}, in much the same way as the (similarly non-renormalisable) chiral perturbation theory of low energy pions \cite{Gasser:1983yg,Weinberg:1978kz,Colangelo:1995np,Bijnens:1998yu,Bijnens:1999hw,Buchler:2003vw}. 

Our initial motivation was to explore the possibility that the renormalisation group (RG) in this context might provide a route to learning something useful about the \emph{non-perturbative} behaviour of quantum gravity. 
In particular, even in a perturbatively non-renormalisable theory, the RG relates the leading UV divergence at each new loop order $\ell$ to one-loop ($\ell=1$) divergences  \cite{Buchler:2003vw}. More physically, it allows us to compute in this way the leading log power $(\ln\mu)^\ell$, of the standard arbitrary RG energy scale $\mu$, at each loop order $\ell$. (These are called chiral logs in pionic perturbation theory \cite{Gasser:1983yg,Weinberg:1978kz,Colangelo:1995np,Bijnens:1998yu,Bijnens:1999hw,Buchler:2003vw}.) If it were possible to use the RG relations to compute these leading terms to arbitrarily high loop order, and resum them, we would get a powerful insight into the UV behaviour of quantum gravity at the {non-perturbative} level. 

In perturbative quantum gravity the leading divergences actually vanish on-shell. They are therefore field reparametrisations, and have no effect on the $S$-matrix. However if we keep in mind that the UV behaviour of the full two-point correlator is characterised by its \emph{off-shell} dependence, we see that these leading divergences and associated powers of $\ln \mu$, could nevertheless be important. For example, after resumming them,  one might find  that the non-perturbative UV behaviour of the two-point correlator, and potentially thus that of quantum gravity more generally, is very different from what one would na\"\i vely conclude order by order in perturbation theory. 

In a non-renormalisable gauge theory, divergences that vanish on the equations of motion (of the quantum fields), are related to modifications of the BRST algebra \cite{Gomis:1995jp} (see also \cite{Dixon:1975si,Voronov:1982cp,Voronov:1982ph,Lavrov:1985ugv,Anselmi:1994zx}).\footnote{Actually this was established only for vanishing background field. We treat the non-vanishing case in sec. \ref{sec:General form of divergences in the Legendre effective action}.} At each loop order the corresponding counterterms modify the BRST algebra in a way that remains consistent with the Zinn-Justin identities \cite{ZinnJustin:1974mc,ZinnJustin:2002ru}.
They do this by generating \emph{canonical} reparametrisations of the antifields (sources for BRST transformations) \cite{Batalin:1981jr,Batalin:1984jr,Batalin:1984ss,Gomis:1994he} and quantum fields. 

On the other hand as we already mentioned, in a generic non-renormalisable theory the RG tells us that the leading  divergences can be expressed recursively in terms of divergences in one-loop diagrams, namely one-loop counterterm diagrams, being those that contain at least one counterterm vertex \cite{Buchler:2003vw}. As we demonstrate in sec. \ref{sec:RGandCME}, these recurrence relations are actually crucial for consistency of the above canonical transformations. Unfortunately for a non-renormalisable theory, the one-loop counterterm diagrams are themselves new and non-trivial at each new loop order, and thus provide a practical obstruction to deriving the leading divergence at arbitrary order. 

Viewed in this light the proposal of ref. \cite{Solodukhin:2020vuw}, would appear to potentially provide a breakthrough. The key idea is to exploit the pole equations that follow from assuming finite generalised $\beta$-functions for the field reparametrisations.  As we will see in sec. \ref{sec:solodukhin}, they imply that the leading divergences at higher loops ($\ell>1$) should actually be computable by recursive differentiation, in particular without computing any more Feynman diagrams. Unfortunately, the proposal is not correct as will become clear in this paper. We spell this out in detail in sec. \ref{sec:solodukhin}. 

One problem with exploring these ideas is that there are effectively no explicit higher-loop off-shell leading divergences in the literature that one can test against. Some purely background field off-shell two-loop $1/\eps^2$ divergences appear in the famous paper ref. \cite{Goroff:1985th}, but unfortunately they contain an error, as pointed out in ref. \cite{Solodukhin:2020vuw}.

All of the above considerations motivated us to compute explicitly (in Feynman -- De Donder gauge and dimensional regularisation) the leading off-shell divergences for the two-point vertex up to two loops, and in particular to draw out their intimate relation to the one-loop counterterm diagrams \cite{Buchler:2003vw} and to canonical transformations in the BRST algebra \cite{Gomis:1995jp}.
Since this necessitates computing, as an intermediate step, the off-shell divergences in one-loop diagrams with three external legs, two of which are quantum, we widened our investigation so as to provide explicit results for all off-shell one-loop divergences with up to three fields.

In fact even for just the graviton one-loop two-point divergence, the complete results do not appear in the literature. Famously, the pure background part appears in ref. \cite{tHooft:1974toh}. The pure quantum part appears in ref. \cite{Capper:1979ej}, \cf also app. \ref{app:comparisons}, and ref. \cite{Kellett:2020mle}.  But to our knowledge the divergence in the mixed quantum background vertex has not appeared before in the literature. These three divergences can be expressed in terms of appropriately defined linearised curvatures. (For the quantum field, this is an accident of Feynman -- De Donder gauge, \cf sec. \ref{sec:onelooptwoptagone}.) However the three expressions are all different (thus not as assumed in ref. \cite{Chase:1982sf}). Although they are all different, they are not independent. Their relation is precisely such that all three are removed by a canonical transformation of the quantum fields (and antifields). 

This may come as a surprise since \textit{a priori} one might expect that a separate reparametrisation of the background metric should also be performed (in fact this is what is assumed and employed in ref. \cite{Solodukhin:2020vuw}). However in sec. \ref{sec:General form of divergences in the Legendre effective action} we show in general that this does not happen. New divergences at each loop order which involve background and quantum fluctuations and do not vanish on the equations of motion, are purely a function of the total metric (that combines background and fluctuation), whilst all other divergences are renormalised by a canonical transformation of the quantum fields and antifields.

We show explicitly that this scenario continues to hold at the three-point level, where now thousands of vertices are divergent. We verify that the divergence in the Gauss-Bonnet topological term \cite{Gibbons:1978ac,Goroff:1985th}  is indeed a function only of the total metric, whilst all other divergences are removed by a canonical transformation on the antifields and quantum fields. 

Then in sec. \ref{sec:twoloops} we use the one-loop counterterm diagrams to derive the leading divergence at two loops in the pure background, pure quantum, and mixed, two-point vertices. At this stage the dependence on the quantum field can no longer be written in terms of linearised curvatures, reflecting the fact that BRST transformations are now modified to the extent that they do not reduce to diffeomorphisms. Nevertheless, taking proper account of non-linearities in the Zinn-Justin equations, we verify again that all these divergences can be removed by a canonical transformation on the antifields and quantum fields.

The structure of the paper is as follows. In sec. \ref{sec:off-shell BRST} we define the BRST transformations for the quantum fluctuation field and ghosts in the presence of a background metric. We develop the formalism that is needed to cope with the fact that BRST invariance is significantly altered in the process of renormalisation. Consistency is maintained by preserving the Zinn-Justin equation \cite{ZinnJustin:1974mc,ZinnJustin:2002ru} \aka CME (Classical Master Equation) \cite{Batalin:1981jr,Batalin:1984jr,Batalin:1984ss}. We work with so-called off-shell BRST and display results in so-called minimal basis, since it provides the most elegant and powerful realisation, 
but in sec. \ref{sec:Minimal basis and comparisons to on-shell BRST} we explain why the calculations themselves are essentially the usual ones. Both the bare action and the Legendre effective action satisfy the Zinn-Justin equation \cite{ZinnJustin:1974mc,ZinnJustin:2002ru} as we review in secs.  \ref{sec:The CME for the bare action} and \ref{sec:CMEs} respectively, but beyond one loop this leads to a tension and this tension is resolved by the RG relations for counterterm diagrams, as we explain in secs. \ref{sec:RGandCME} and \ref{sec:RG}. 

New divergences are invariant under the total classical BRST charge $s_0$ which incorporates not only the BRST transformations but also the action of the Koszul-Tate operator. Taking into account the presence of the background metric, their properties are developed in sec. \ref{sec:Properties of the total classical BRST charge}. Since $s_0$ is nilpotent, solutions are classified according to its cohomology. As we recall in sec. \ref{sec:canonsecond}, those solutions that are $s_0$-exact are first order canonical transformations of the CME. At two loops we need also the canonical transformations to second order and their relation to the perturbatively expanded CME. This is derived in sec. \ref{sec:canonsecond}. Then in sec. \ref{sec:General form of divergences in the Legendre effective action} we derive the general solution for $s_0$-closed divergences. We show that cohomologically non-trivial solutions can be taken to be functions of only the total metric, with the rest being $s_0$-exact, in particular there are no separate purely background metric divergences. 

As already mentioned, in sec. \ref{sec:Loops} we compute for the first time many off-shell counterterms that appear up to two loops, and use them to verify all these properties. In this way also we provide a concrete example of how the BRST transformations get appreciably modified by loop  corrections. In sec. \ref{sec:solodukhin} we investigate the proposal for generalised beta-functions for field reparametrisations. We start by assuming as in the original proposal that it is the background metric that should be reparametrised and then, given the results of this paper, put forward a more natural scenario where the beta functions are built on the canonical transformations. Unfortunately neither of these ideas lead to finite beta functions, and we explain why they cannot.  Finally in sec. \ref{sec:conclusions} we draw our conclusions.

\section{BRST in perturbative quantum gravity and its renormalisation}
\label{sec:off-shell BRST}

In this section we first set up the BRST framework that we will use, and then develop its properties. Along the way we make a number of new observations. 
In particular we will see in sec. \ref{sec:RGandCME} that RG invariance is actually essential to ensure that the BRST symmetry can be renormalised successfully, whilst in sec. \ref{sec:General form of divergences in the Legendre effective action} we prove the absence of a separate background field divergence in new divergences at each loop order.

\subsection{The CME for the bare action}
\label{sec:The CME for the bare action}

In a perturbative setting we work with a quantum, \aka fluctuation, field $h_{\mu\nu}$. This field is defined by our choice of expansion of the (total) metric $g_{\mu\nu}$ around a background metric $\Bar{g}_{\mu\nu}$. In this paper we simply set 
\begin{equation}
\label{background expansion}
    g_{\mu\nu} = \Bar{g}_{\mu\nu} + \kappa\, h_{\mu\nu} \,.
\end{equation}
where $\kappa=\sqrt{32\pi G}$ is the natural expansion parameter, $G$ being Newton's gravitational constant. We are interested in off-shell divergences, and their value depends on the choice of expansion. Using the above allows us to compare with previous results in the literature \cite{tHooft:1974toh,Capper:1979ej,Goroff:1985th,Kellett:2020mle}.

We will work with so-called \emph{off-shell} BRST \cite{Becchi:1974xu,Becchi:1974md,Becchi:1975nq,Tyutin:1975qk}.
In this way we can fully exploit BRST invariance at every step, and keep track of how it changes under quantum corrections.
Although we only actually need the Zinn-Justin equation \cite{ZinnJustin:1974mc,ZinnJustin:2002ru} for this, it is convenient to phrase the calculation in terms of the Batalin-Vilkovisky formalism \cite{Batalin:1981jr,Batalin:1984jr,Batalin:1984ss}, employing known identities for the antibracket \cite{Batalin:1981jr,Batalin:1984jr,Batalin:1984ss,Gomis:1994he}:
\begin{equation}
\label{BV antibracket}
    \left( X , Y \right) = \frac{\partial_r X}{\partial \phi^A} \frac{\partial_l Y}{\partial \phi^*_A} - \frac{\partial_r X}{\partial \phi^*_A} \frac{\partial_l Y}{\partial \phi^A} \,,
\end{equation}
where $X$ and $Y$ are two functionals, $\phi^A$ are the quantum fields (including ghosts $c^\mu$) and $\phi^*_A$ are the antifields (opposite statistics sources for the BRST transformations $Q\phi^A$ of the corresponding fields), and we are here employing compact DeWitt notation (so Einstein summation over the capital indices indicates both summation over Lorentz indices and integration over spacetime).
As we will see, the resulting framework allows calculations that are no more onerous than standard ones employing only on-shell BRST invariance \cite{Morris:2018axr,first,Kellett:2020mle}. Furthermore, we can then display the results more compactly by using the so-called minimal basis \cite{Batalin:1981jr,Morris:2018axr,first,Kellett:2020mle}. 

We choose the bare action $S[\phi,\phi^*]$ to include these sources. It will be made up of the classical action $S_0$ plus a series of local counterterms $S_\ell$ chosen to cancel the divergences that appear at each loop order $\ell$, whilst introducing the new renormalized couplings (\cf sec. \ref{sec:RG} \cite{Buchler:2003vw}) which, because they run with $\mu$, must also be introduced at that order:
\be \label{bare}  S=S_0 + \hbar S_1 + \hbar^2S_2 +\cdots\,. \ee
By including the sources $\phi^*$ we will additionally incorporate the counterterms necessary to render finite the BRST transformations \cite{ZinnJustin:1974mc,ZinnJustin:2002ru}.

At the classical level the bare action is thus given by
\begin{equation}
\label{minimal classical action}
    S_0 = -\int_x \left\{ \frac{2}{\kappa^2} \sqrt{g} R + \left( Q h_{\mu\nu} \right) h^{*\mu\nu} + \left( Q c^\mu \right) c^*_\mu\right\} \,.
\end{equation}
The first term is the Einstein-Hilbert action in Euclidean signature. In this paper we take the cosmological constant to vanish. At the perturbative level, divergences do not force its introduction, so working in this simplified setting is consistent. The integral is over 
\be
\label{dimreg}
d=4-2\eps 
\ee 
dimensional spacetime (we will be using dimensional regularisation). Our conventions for curvatures are 
$R_{\mu\nu}=R^\alpha_{\ \mu\alpha\nu}$, and $[\nabla_\mu,\nabla_\nu]v^\lambda = R_{\mu\nu\phantom{\lambda}\sigma}^{\phantom{\mu\nu}\lambda}v^\sigma$.

For convenience, we choose to define the antifields to have indices in the position shown and to transform as tensor densities of weight $-1$ so that no metric is required above for these terms. Also for convenience, a minus sign is included so that none appears in the identity:
\be \label{QPhi} Q\phi^A = (S_0,\phi^A)\,. \ee
Note that this defines our charges to act from the left. 
Classically, the BRST charge $Q$ can be defined in terms of the Lie derivative along $\kappa c^\mu$:
\beal 
\label{QHLie}
 Q h_{\mu\nu} &= \frac1\kappa Q g_{\mu\nu} = \mathcal{L}_c g_{\mu\nu} = 2 \partial_{(\mu} c^\alpha g_{\nu)\alpha} + c^\alpha \partial_\alpha g_{\mu\nu}\,,\\ 
 \label{QcLie}
 Q c^\mu &= \frac{\kappa}{2} \mathcal{L}_c c^\mu = \kappa c^\nu \partial_\nu c^\mu \,.
\eeal 
Its nilpotence ($Q^2=0$), and diffeomorphism invariance of the Einstein-Hilbert action, implies 
\be 0 = QS_0= Q\phi^A\frac{\partial_lS_0}{\partial\phi^A} = -\frac{\partial_r S_0}{\partial\phi^*_A}\frac{\partial_lS_0}{\partial\phi^A}=\frac12(S_0,S_0)\,,\ee
and thus that the classical bare action $S=S_0$ satisfies the so-called CME (Classical Master Equation) \cite{Batalin:1981jr,Batalin:1984jr,Batalin:1984ss}, \aka Zinn-Justin equation \cite{ZinnJustin:1974mc,ZinnJustin:2002ru}. Once we consider quantum corrections, it is not the BRST transformations (\ref{QHLie},\ref{QcLie}) that we can preserve but only the CME, \ie we will ensure that to any loop order $\ell$ the bare action satisfies:
\be \label{CME}  (S,S) = 0\,. \ee

\subsection{Canonical transformation to gauge fixed basis}
\label{sec:Canonical transformation to gauge fixed basis}

To get the gauge fixed version, we need to work in the so-called extended basis, which introduces a new field and antifield over and above what we already have (the so-called minimal basis) \cite{Batalin:1981jr,Batalin:1984jr,Batalin:1984ss}:
\begin{equation}
\label{extended action}
    S^{(ext)} = S + \int_x \left\{ \frac{1}{2 \alpha}\sqrt{\Bar{g}} \Bar{g}_{\mu\nu} b^\mu b^\nu + i b^\mu \Bar{c}^*_\mu \right\} \,,
\end{equation}
where $\alpha$ is the gauge parameter, $b^\mu$ is a bosonic auxiliary field, and $\Bar{c}^*_\mu$ sources the BRST transformation for the antighost. From \eqref{QPhi} we have $Q\Bar{c}^\mu=-ib^\mu$ and $Qb^\mu=0$. Trivially, the CME and $Q^2=0$ continue to hold. The next step is to introduce a suitable gauge fixing fermion $\Psi[\phi]$. In the Batalin-Vilkovisky treatment this is used to eliminate the antifields \cite{Batalin:1981jr,Batalin:1984jr,Batalin:1984ss,Gomis:1994he}. We keep them however, because of their crucial r\^ole in renormalisation, and in particular in the Zinn-Justin identities, and instead get the same effect by performing an exact canonical transformation \cite{Gomis:1994he}
\begin{align}
    \check{\phi}^A &= \frac{\partial_l}{\partial \check{\phi}^*_A} \K[\phi,\check{\phi}^*] \,,\nn \\
    \phi^*_A &= \frac{\partial_r}{\partial \phi^A} \K[\phi,\check{\phi}^*] \,,\label{canon}
\end{align}
from the above gauge invariant (g.i.) basis $\{\phi,\phi^*\}$, to a gauge fixed (g.f.) basis $\{\check{\phi},\check{\phi}^*\}$, setting \cite{Morris:2018axr,first,Kellett:2020mle}
\begin{equation}
\label{canonpsi}
    \K = \check{\phi}^*_A \phi^A - \Psi[\phi] \,.
\end{equation}
The advantage of employing a canonical transformation is that by definition it leaves the antibracket invariant and thus in the new basis the CME continues to hold. We choose
\begin{equation}
\label{BF gauge fixing fermion}
    \Psi = \int_x \sqrt{\Bar{g}} F_\mu \Bar{c}^\mu \,,
\end{equation}
and choose DeDonder gauge by setting $F_\mu$ to 
\begin{align}
    F_\mu &= \Bar{\nabla}_\nu h^\nu{}_\mu - \Bar{\nabla}_\mu \ph \,, \label{F}\\
    \ph &= \frac{1}{2} h^\mu{}_\mu = \frac{1}{2} \Bar{g}^{\mu\nu} h_{\mu\nu}\label{phi} \,.
\end{align}
This breaks the diffeomorphism invariance as realised through the total metric $g_{\mu\nu}$ (as required) but leaves it realised as ``background diffeomorphism'' invariance, using the background metric $\Bar{g}_{\mu\nu}$. From here on we raise and lower indices using the background metric, unless explicitly mentioned otherwise, 
and employ the background covariant derivative $\Bar{\nabla}_\mu$ (using the background metric Levi-Civita connection). As is well known, we can put a connection in for free in Lie derivatives, so to make background diffeomorphism invariance manifest in (\ref{QHLie},\ref{QcLie}) we can write the classical BRST transformations (in minimal basis) instead as
\beal 
    Q h_{\mu\nu} &= 2  \Bar{\nabla}_{(\mu} c^\alpha g_{\nu)\alpha} +  c^\alpha \Bar{\nabla}_\alpha g_{\mu\nu}  = 
    2\Bar{\nabla}_{(\mu} c_{\nu)} +2\kappa \Bar{\nabla}_{(\mu}c^\alpha h_{\nu)\alpha} +\kappa c^\alpha\Bar{\nabla}_\alpha h_{\mu\nu}\,,\nn\\
    Q c^\mu &= \kappa c^\nu \Bar{\nabla}_\nu c^\mu \,.\label{Qbackg}
\eeal
Applying the canonical transformation we see that only the following antifields change:
\begin{align}
    h^{*\mu\nu} \big|_{g.f.} &= h^{*\mu\nu} \big|_{g.i.} - \sqrt{\Bar{g}} \left( \Bar{\nabla}^{(\mu} \Bar{c}^{\nu)} - \frac{1}{2} \Bar{\nabla}_\alpha \Bar{c}^\alpha \Bar{g}^{\mu\nu} \right) \,,\label{gf to gi 1}\\
    \Bar{c}^{*\mu} \big|_{g.f.} &= \Bar{c}^{*\mu} \big|_{g.i.} + \sqrt{\Bar{g}} F^\mu \,, \label{g.i. basis}
\end{align}
thus mapping the extended action \eqref{extended action} at the classical level to
\begin{equation}
    S_0^{(ext)} \big|_{g.f.} = S_0 + \int_x \left\{ \frac{1}{2 \alpha}\sqrt{\Bar{g}} \Bar{g}_{\mu\nu} b^\mu b^\nu - i \sqrt{\Bar{g}} F_\mu b^\mu  + i b^\mu \Bar{c}^*_\mu \right\} 
    + \int_x \sqrt{\Bar{g}}  \left( \Bar{\nabla}^{(\mu} \Bar{c}^{\nu)} - \frac{1}{2} \Bar{\nabla}_\alpha \Bar{c}^\alpha \Bar{g}^{\mu\nu} \right) Q h_{\mu\nu}\,. \label{g.f. action}
\end{equation}
The first term is \eqref{minimal classical action}, the classical action in minimal basis, and the last term is the usual ghost action (in DeDonder gauge). 
The middle term is purely quadratic in $b^\mu$. We could thus integrate it out. Dropping the $\Bar{c}^*_\mu$, the integrand is:
\begin{equation}
\label{int out}
    \frac{\sqrt{\Bar{g}}}{2\alpha}(b_\mu-iF_\mu)^2+\frac{\alpha}{2}\sqrt{\Bar{g}}F^\mu F_\mu \,.
\end{equation}
The $b_\mu$ integral over the first term vanishes in dimensional regularization, whilst 
the second term is the standard gauge fixing term. In fact this is now the textbook \textit{on-shell BRST} treatment. The action $S_0$ is still BRST invariant if we now set $Q \Bar{c}^\mu = \alpha F^\mu$. But this is not quite as powerful because $Q^2 \Bar{c}^\mu = \alpha Q F^\mu$, only vanishes on shell ($Q F_\mu = 0$ is the $\Bar{c}$ equation of motion). For this reason we keep $b^\mu$ and stick with this off-shell BRST treatment. 

Since we will be working with a perturbative expansion over quantum fields and antifields, we may as well treat the background metric perturbatively also. Following \eqref{background expansion}, we write:
\begin{equation}
\label{gback expansion}
    \bar{g}_{\mu\nu} = \delta_{\mu\nu} +\kappa \bar{h}_{\mu\nu}\,,\qquad\implies\qquad g_{\mu\nu} = \delta_{\mu\nu} + \kappa\,  h_{\mu\nu} +\kappa \bar{h}_{\mu\nu}\,.
\end{equation}
At this stage we can invert the terms bilinear in the quantum fields to get the propagators. 
For general $\alpha$ gauge see \eg ref. \cite{Morris:2018axr}. We will use Feynman gauge, $\alpha=2$, which gives the simplest propagators. Once again, the coefficients of off-shell divergences depend on these choices. By using Feynman DeDonder gauge we make the same choices as in older works \cite{tHooft:1974toh,Capper:1979ej,Goroff:1985th,Kellett:2020mle} and can thus compare our results. Writing 
\be 
\phi^A(x) = \int \!\frac{d^dp}{(2\pi)^d}\, \text{e}^{-i p\cdot x}\, \phi^A(p)\,,
\ee
we have:
\beal
\label{HHalph}
\langle h_{\mu\nu}(p)\,h_{\alpha\beta}(-p)\rangle &= \frac{\delta_{\mu(\alpha}\delta_{\beta)\nu}}{p^2}
-\frac1{d-2}\frac{\delta_{\mu\nu}\delta_{\alpha\beta}}{p^2}\,, \\
\langle b_\mu(p) \,h_{\alpha\beta}(-p)\rangle &= -\langle h_{\alpha\beta}(p)\,b_\mu(-p)\rangle  = 2\, \delta_{\mu (\alpha} p_{\beta)}/{p^2}\,,\\
\langle b_\mu(p)\,b_\nu(-p)\rangle &= 0\,,\\
\langle c_\mu(p)\, \bar{c}_\nu(-p)\rangle &= -\langle \bar{c}_\mu(p)\, c_\nu(-p) \rangle =  \delta_{\mu\nu}/{p^2}\,.
\eeal

\subsection{Minimal basis and comparisons to on-shell BRST}
\label{sec:Minimal basis and comparisons to on-shell BRST}

We will be computing quantum corrections to the one-particle irreducible, \aka Legendre,  effective action $\Gamma$. Since we have an auxiliary field $b^\mu$ and the extra propagator $\langle b_\mu h_{\alpha\beta} \rangle$, at first sight this formalism complicates the computation and cannot be directly compared to earlier results using on-shell BRST \cite{tHooft:1974toh,Capper:1979ej,Goroff:1985th}. However this is not the case. 

\begin{figure}[ht]
\centering
$$
\begin{gathered}
\begin{tikzpicture}
\begin{feynman}
\vertex  (a) at (0,0);
\vertex (b) at (-0.4,0.4);
\vertex (c) at (0.4,0.4);
\vertex (d) at (0.3,-0.3);
\vertex (e) at (-0.3,-0.3);
\vertex (d1) at (0.6,-0.6);
\vertex (e1) at (-0.6,-0.6);

\diagram*{
(a) -- [photon] (b),
(a) -- [photon] (c),
(a) -- [plain] (d1),
(a) -- [dashed] (e),
(b) -- [ghost, quarter left] (c),
(e) -- [plain] (e1),
};
\end{feynman}
\end{tikzpicture}
\end{gathered}
\quad
\begin{gathered}
\begin{tikzpicture}
\begin{feynman}
\vertex  (a) at (0,0);
\vertex (b) at (-0.4,0.4);
\vertex (c) at (0.4,0.4);
\vertex (d) at (0.3,-0.3);
\vertex (e) at (-0.3,-0.3);
\vertex (d1) at (0.6,-0.6);
\vertex (e1) at (-0.6,-0.6);

\diagram*{
(a) -- [photon] (b),
(a) -- [photon] (c),
(a) -- [dashed] (d),
(a) -- [dashed] (e),
(b) -- [ghost, quarter left] (c),
(e) -- [plain] (e1),
(d) -- [plain] (d1),
};
\end{feynman}
\end{tikzpicture}
\end{gathered}
\quad
\begin{gathered}
\begin{tikzpicture}
\begin{feynman}
\vertex  (a) at (0,0);
\vertex (b) at (0,0.5);
\vertex (c) at (-0.5,0.4);
\vertex (d) at (0.4,0);
\vertex (e) at (-0.3,-0.3);
\vertex (e1) at (-0.6,-0.6);
\vertex (a1) at (1,0);
\vertex (b1) at (1,0.5);
\vertex (c1) at (1.5,0.4);
\vertex (c2) at (1.6,-0.6);
;

\diagram*{
(a) -- [photon] (b),
(a) -- [photon] (c),
(a) -- [plain] (d),
(a) -- [dashed] (e),
(e) -- [plain] (e1),
(b) -- [ghost, quarter right] (c),
(d) -- [dashed] (a1),
(a1) -- [photon] (b1),
(a1) -- [photon] (c1),
(b1) -- [ghost, quarter left] (c1),
(a1) -- [plain] (c2),
};
\end{feynman}
\end{tikzpicture}
\end{gathered}
$$
 \caption{Examples that illustrate that one-particle irreducible Feynman diagrams involving $b$ interactions with an unspecified number of external background metric $\bar{h}$ legs (fan of wavy lines), arise by starting with an internal $h$ (solid line) propagating into $b$ (dashed line) and eventually back to $h$. These implement in diagrammatic language the effect \eqref{int out} of integrating out the $b$ field.}
 \label{fig:intoutb}
\end{figure}
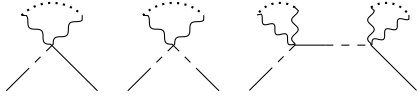

First note that the $h$ propagator \eqref{HHalph} is the same as in the usual treatment. (This is actually guaranteed in any gauge, but we omit the proof.) Setting $\bar{h}_{\mu\nu}=0$ for the moment, we note that the interaction terms (\ie with three or more fields) in \eqref{g.f. action} do not contain $b^\mu$ or $\Bar{c}^*_{\mu}$. Feynman diagram contributions to $\Gamma$ therefore have the same property and coincide with those computed in the usual (on-shell BRST) treatment. Switching back on the background metric, we do now have interactions involving the background metric and either $b^2$, or $b$ and $h$. However it is not possible then to draw one-particle irreducible diagrams with external $b$-field legs. The interactions only contribute in diagrams by having $h$ propagate to $b$ and back again, see fig. \ref{fig:intoutb}, and the net effect of including all these corrections is to incorporate in diagrammatic language the result of integrating out $b$. Thus these Feynman diagrams simply reproduce the corrections we get from the second term in \eqref{int out}, \ie the standard gauge fixing term. So we see that we can continue to ignore $b^\mu$ and $\Bar{c}^*_{\mu}$ provided we include the interactions from the standard gauge fixing term. Furthermore, we get in this way the same results as the standard treatment. 

Next note that the corrections only depend on $\Bar{c}^\mu$ through the combination on the right-hand side of \eqref{gf to gi 1}. This means that we can shift back to g.i. basis after computing loop contributions to $\Gamma$, the only dependence on $b$ and $\Bar{c}^*$ then being as in the extended action \eqref{extended action}. Furthermore we can then display results in minimal basis by removing the $b$ and $\Bar{c}^*$ terms. 

This all means that we can construct $\Gamma$ order by order in the minimal basis, never needing $b$ or $\Bar{c}^*$. To do so we shift $h^{*\mu\nu} \big|_{g.i.}$ to $h^{*\mu\nu} \big|_{g.f.}$ in interactions and use the $\langle h_{\alpha\beta} h_{\mu\nu} \rangle$ and $\langle c_\mu \Bar{c}_\nu \rangle$ propagators and include the interaction vertices from the standard gauge fixing term \eqref{int out} as appropriate, and afterwards shift back to g.i. basis \cite{first,Kellett:2020mle}. Of course this does not mean that off-shell quantum corrections are independent of our choice of gauge. However the results are sometimes much simpler when cast back in (minimal) g.i. basis in this way, which is why we use it in this paper.

\subsection{The CME for the Legendre effective action}
\label{sec:CMEs}

Since the BRST transformations \eqref{Qbackg}, or (\ref{QHLie},\ref{QcLie}), involve products of fields at the same spacetime point, they are not preserved under renormalisation. Order by order in the loop expansion not only must the action be modified, but also the BRST transformations themselves, and since the theory is non-renormalisable, the changes involve in fact an infinite series in powers of the fields and antifields. The Zinn-Justin equation  \cite{ZinnJustin:1974mc,ZinnJustin:2002ru,Batalin:1981jr,Batalin:1984jr} can keep track of all this. We start with the fact that the partition function
\begin{equation}
\label{partnfn}
    \mathcal{Z} \equiv \mathcal{Z}[J,\phi^*] = \int\!\! \mathcal{D} \phi\, \mathrm{e}^{-S[\phi,\phi^*] + \phi^A J_A} \,,
\end{equation}
satisfies the identity 
\be\label{Ward}\frac{\partial_r \mathcal{Z}}{\partial \phi^*_A} J_A =0\,.\ee 
To prove this at the classical level it is sufficient to use the fact that $QS_0=0$, assuming invariance of the measure:
\be
\label{Wardcl}
    0 = \int\!\! \mathcal{D}\phi\, Q \!\left( \mathrm{e}^{-S_0+\phi^A J_A} \right) 
      = \int\!\! \mathcal{D}\phi\, \mathrm{e}^{-S_0+\phi^A J_A} \left( Q \phi^A \right) J_A 
      = - \frac{\partial_r \mathcal{Z}}{\partial \phi^*_A} J_A \,.
\ee
But at the quantum level we need to derive it via preservation of the CME, \eqref{CME}: 
\be
\label{Wardq}
0  = \int\!\! \mathcal{D}\phi\, \frac{\partial_l}{\partial\phi^A}\frac{\partial_r}{\partial\phi^*_A}\, \mathrm{e}^{-S[\phi,\phi^*] + \phi^A J_A}
= -\int\!\! \mathcal{D}\phi\, \left\{J_A\frac{\partial_rS}{\partial\phi^*_A}+ \frac12(S,S)+\frac{\partial_l}{\partial\phi^A}\frac{\partial_r}{\partial\phi^*_A}S\right\} \, \mathrm{e}^{-S[\phi,\phi^*] + \phi^A J_A}\,.
\ee
Here the first equality follows because it is an integral of a total derivative. After rearranging the result using the statistics of the (anti)fields, we get the three terms inside the braces. The first term gives the required identity, the second term vanishes by the CME, whilst the third term is the Batalin-Vilkovisky measure term \cite{Batalin:1981jr,Batalin:1984jr,Batalin:1984ss}. In general we need to take this into account (giving the Quantum Master Equation) \cite{Batalin:1981jr,Batalin:1984jr,Batalin:1984ss,first,Kellett:2020mle,Morris:2018axr} however, since $S$ is local, this term always contains $\delta(x)|_{x=0}$ or its space-time derivatives. These vanish in dimensional regularisation. Therefore in this paper
we can discard the measure term.

Introducing the generator $W[J,\phi^*]$ of connected diagrams, through $\mathcal{Z} = \mathrm{e}^W$, 
we define the Legendre effective action in the usual way:
\be\label{Gamma} \Gamma[\Phi,\Phi^*] = - W + \Phi^A J_A\,,\qquad \Phi^A = \frac{\partial_r W}{\partial J_A}\,,\quad J_A = 
\frac{\partial_l\Gamma}{\partial\Phi^A}\,, \ee
where $\Phi^A$ is  the so-called classical field, and we have renamed $\phi^*_A\equiv\Phi^*_A$ just because it looks better. Then by standard manipulations \eqref{Ward} turns into the Zinn-Justin equation:
\begin{equation}
\label{CMEGamma}
     \left( \Gamma , \Gamma \right) = 0 \,,
\end{equation}
\ie again the CME \eqref{CME}, now applied to $\Gamma[\Phi,\Phi^*]$, the antibracket taking the same form as \eqref{BV antibracket} but with $\{\phi,\phi^*\}$ replaced with $\{\Phi,\Phi^*\}$.


The Legendre effective action 
\be \Gamma = \Gamma_0+\hbar\Gamma_1+\hbar^2\Gamma_2+\cdots\,,\ee
is built up recursively, where $\Gamma_\ell$ is the $\ell$-loop contribution, starting with $\Gamma_0 = S_0$, the classical bare action.
The logic now is to introduce at each new loop order $\ell$, a local counterterm action $S_\ell$ to the bare action in order to cancel the divergences $\Gamma_\ell |_\infty$ that arise in $\Gamma_\ell$, leaving behind an arbitrary finite part which is parametrised by the new renormalized couplings that appear at this order. Provided we introduce $S_\ell$ in such a way as to preserve $(S,S)=0$ we also have that $(\Gamma,\Gamma)=0$ is satisfied. However, although
both the bare action $S$ and the Legendre effective action $\Gamma$ satisfy the CME, the CME plays a different r\^ole in each case so that it is in fact not trivial that the two are consistent beyond one loop.  As we will see what makes them nevertheless consistent is the RG. 

\subsection{How the RG is needed for consistent solutions to both versions of the CME}
\label{sec:RGandCME}

Expanding the CME \eqref{CMEGamma} for $\Gamma$, we see that the one-loop contribution satisfies $(\Gamma_0,\Gamma_1)=0$. It is useful to define the total classical BRST charge $s_0$ acting on any functional $X$ as
\be 
\label{s}
s_0 X = (S_0,X)\,,
\ee
which thus acts also on antifields (see sec. \ref{sec:Properties of the total classical BRST charge}),
then the one-loop BRST identity is simply $s_0\Gamma_1=0$. Since dimensional regularisation is a gauge invariant regulator, the infinite part, which at one loop is proportional to a single pole, $\propto1/\eps$, also satisfies this identity, \ie
\be s_0\,\Gamma_{1/1}[\Phi,\Phi^*]=0\,. \label{oneloopdivclosed}\ee
(We label terms proportional to divergences $1/\eps^k$, by appending $/k$ to the subscript.)

It is simplest for our purposes to now consider the identity satisfied by the two-loop contribution, $\Gamma_2$, before any renormalisation. 
From the CME \eqref{CMEGamma} we see that it satisfies
\be s_0\Gamma_2 = -\frac12(\Gamma_1,\Gamma_1)\,. \label{twoloopCME}\ee 
In particular this implies for the double-pole divergence:
\be\label{divtwoloop} s_0\,\Gamma_{2/2} = -\frac12\left(\Gamma_{1/1},\Gamma_{1/1}\right)\,. \ee
Given that the right hand side does not vanish, this is a non-trivial relation between the $1/\eps^2$ divergences at two loops and the $1/\eps$ divergences at one loop.

Now we consider the process of renormalisation. At one loop, if we add a counterterm action $S_1$, then in order to preserve the CME \eqref{CME} for $S$, we find in the same way that $S_1$ must be chosen so that it is also annihilated by the total classical BRST charge:  
\be 
\label{SoneCME}
s_0S_1=0\,.
\ee 
Since the one-loop divergence is local we can then render the one-loop result finite by setting 
\be\label{ctoneloopstructure} S_1 = - \Gamma_{1/1}[\phi,\phi^*] + S_{c_1}[\phi,\phi^*]\,,\ee
where the finite remainder $S_{c_1}$ contains the new renormalised couplings $c^j_1(\mu)$ that appear at one loop, \cf sec. \ref{sec:RG}, in particular they are needed for the curvature-squared terms but also for antifield vertices, see secs. \ref{sec:onelooptwopt}, \ref{sec:threepoints}. Clearly we must also have $s_0S_{c_1}=0$. 

Expanding the CME \eqref{CME} to $O(\hbar^2)$, we find of course an algebraically identical formula to \eqref{twoloopCME}, \eqref{divtwoloop}:
\be 
\label{StwoCME}
s_0S_2=-\frac12(S_1,S_1)\,.
\ee
This must be satisfied by the counterterm action $S_2$.
It relates the $1/\eps^2$ divergence in this two-loop counterterm to the $1/\eps$ divergence in the one-loop counterterms. Then by \eqref{ctoneloopstructure}, we see that the $1/\eps^2$ divergence on the right hand side is precisely the same as in the $\Gamma$ identity \eqref{divtwoloop}. But this is in apparent  contradiction with the fact that $S_2$ \emph{must cancel} the divergence in $\Gamma_2$. In particular the latter implies that $s_0(S_2+\Gamma_2)$ must be finite.

The resolution is that, once we add the one-loop counterterm from $S_1$ to the bare action, at $O(\hbar^2)$ we also have one-loop counterterm diagrams from one-loop diagrams $\Gamma_1[S_1]$ with one $S_1$ vertex inserted (as illustrated in fig. \ref{fig:twopointcountertermdiags} of sec. \ref{sec:twoloops}). The two-loop divergence in \eqref{divtwoloop} comes from diagrams containing only tree level vertices. It must be that the $1/\eps^2$ contribution from the one-loop counterterm diagrams,  is in fact precisely right to flip the sign so that in full the double-pole part satisfies
\be\label{consistency} s_0\,(\Gamma_{2/2}+\Gamma_{1/2}[S_1]) = +\frac12\left(\Gamma_{1/1},\Gamma_{1/1}\right)\,. \ee
As we will see in the next subsection, RG invariance tells us that we have the relation 
\be\label{RGdoublepole}\Gamma_{1/2}[S_1]=-2\Gamma_{2/2}\,,\ee 
and thus for the full double-pole contribution, $\Gamma_{2/2}+\Gamma_{1/2}[S_1]= -\Gamma_{2/2}$, we indeed have the required change of sign (even before the application of $s_0$).
We see therefore that the RG relations are responsible for restoring consistency between the two versions of the CME.  

Although the relations above constrain the form of the double-pole divergences, we still have to compute some Feynman integrals to determine them. Nevertheless we can simplify the process by exchanging the genuinely two-loop diagrams for one-loop counterterm diagrams. The corresponding double-pole counterterm action will automatically satisfy the constraint \eqref{StwoCME}. This latter constraint does not uniquely determine $S_2$ since it is invariant under adding a piece, $S'_2$, provided it is annihilated by the total classical BRST charge: $s_0S'_2=0$. Since this constraint is linear homogeneous, $S'_2$ has finite remainders parametrised by new two-loop couplings $c_2^j(\mu)$. 

We finish this section with some comments about the two-loop \emph{single-pole} divergences. Firstly note that, before adding the one-loop counterterm diagrams, the two-loop single-pole divergences are actually non-local. Indeed, this must be the case since the right hand side of \eqref{twoloopCME} has such non-local divergences in the antibracket contribution $(finite,\Gamma_{1/1})$, where we have written $\Gamma_1=\Gamma_{1/1}+ finite$, and recognised that the finite part is non-local. On adding the counterterm diagrams, the same RG invariance identity that resolves the above putative puzzle, is also responsible for eliminating the non-local divergences  (see the argument of Chase \cite{Chase:1982sf}, which we review in the next subsection). In a similar vein, the two-loop counterterm action $S_2$ has single-pole divergences that depend on the one-loop couplings $c^j_1$, as it must in order to renormalise the $\Gamma_{1/1}[S_1]$ contribution. The fact that $S_2$ must have dependence on $c^j_1$ can also be seen through \eqref{ctoneloopstructure} and the two-loop CME relation, \eqref{StwoCME}. These two constraints must again be related through similar RG identities.

Finally note that there are two-loop single-pole divergences that are not fixed by the RG or by the CME. These will include the famous Goroff and Sagnotti term  \eqref{GS}, but also further terms that vanish on the equations of motion. 
Renormalising them requires new counterterms whose finite remainder introduces further two-loop renormalised couplings $c_2^j(\mu)$. As before, from \eqref{StwoCME} we see that this new part $S'_2$ must be chosen so that it is annihilated by the total classical BRST charge: $s_0S'_2=0$. Thus despite the fact that BRST invariance is significantly altered by the quantum corrections, a central r\^ole is played, order by order in the loop expansion, by the total classical BRST charge $s_0$. We will develop the properties of $s_0$ in sec. \ref{sec:Properties of the total classical BRST charge}.

\subsection{Relating counterterms via the RG}
\label{sec:RG}

Adapting ref. \cite{Buchler:2003vw} to quantum gravity, we prove the RG relation \eqref{RGdoublepole}, which was used in the previous subsection to demonstrate consistency at two loops of the two r\^oles for the CME. 
This key equation relates the double-pole $\Gamma_{1/2}[S_1]$ from the one-loop counterterm diagrams, to the double-pole $\Gamma_{2/2}$ generated by two-loop diagrams using only tree-level vertices. In this subsection, we also review the alternative proof in ref. \cite{Chase:1982sf} for this relation. Rearranging \eqref{RGdoublepole} we see that it implies that the $1/\eps^2$ part of the two-loop counterterm is $-1/2$ times the $1/\eps^2$ pole in the one-loop counterterm diagrams:
\be\label{countertermkey} S_{2/2} = -\left(\Gamma_{2/2}+\Gamma_{1/2}[S_1]\right) = -\frac12\Gamma_{1/2}[S_1]\,.\ee
It is this form that falls out most naturally from the RG analysis, and it is also this form that we use in sec. \ref{sec:twoloops} to compute the $1/\eps^2$ divergence in the two-loop graviton self-energy.

To adapt \cite{Buchler:2003vw}, it proves convenient to absorb Newton's constant into the operators so that the $O(\hbar^0)$ (\ie classical) bare action has pure fluctuation field vertices ($n\ge2$):
\be 
\label{classical}
\cO_{0\,i}\sim\kappa^{n-2}h^np^2\,.
\ee
The numerical subscript on $\cO$ refers to $\hbar$ order \cite{Buchler:2003vw}, and here we are just counting the number of instances of the fluctuation field $h_{\mu\nu}$, $\kappa$ and momentum $p$, where the latter stands for any momentum  (or spacetime derivative) in the vertex, in order to track their dimensions and motivate the formulae below.  Working with pure $h_{\mu\nu}$ vertices will be sufficient to derive \eqref{countertermkey} in this case. Then we will justify why it is clear that \eqref{countertermkey} continues to hold when the background, ghosts and antifields are included. 

In $d=4-2\eps$ dimensions, the mass dimensions are $[h] = -[\kappa] = 1-\eps$.
\textit{A priori} both $\kappa$ and the fluctuation field should be taken to be bare, in the expectation that they will have a divergent expansion in renormalised quantities, but the divergences that are generated involve ever greater powers of momentum, so the vertices in \eqref{classical} are never reproduced and thus neither $\kappa$ nor $h$ require renormalisation. The classical bare action is therefore being written as 
\be 
\label{bareschematic}
S_0 = \Gamma_0 = \int_x c^i_0\,\cO_{0\,i}\,.
\ee 
The $c^i_0$ are the classical couplings with $\kappa$ factored out. They are fixed up to choice of expansion of the metric, choice of gauge fixing, and the value of the cosmological constant if there is one. As mentioned below \eqref{minimal classical action}, in this paper we set the cosmological constant to zero. 

The divergent one-loop quantum corrections then take the form ($H$ is the vacuum expectation value of $h$):
\be 
\label{oneloopdiv}
\Gamma_{1/1} \sim \frac1{\eps} \kappa^n H^np^{4-2\eps}\,,
\ee
\ie in terms of counting overall powers there is an extra factor of $\kappa^2p^{2-2\eps}$. To renormalise we thus have to add to the bare action the local action \eqref{ctoneloopstructure}:
\be 
\label{oneloop}
S_1 = \mu^{-2\eps}\int_x \left\{c^i_1\,\cO_{1\,i} +\frac1\eps a^i_{1/1}\,\cO_{1\,i}\right\}\,,
\ee
where the second set are the counterterms $-\Gamma_{1/1}$, and the first set is the expansion of $S_{c_1}$ and contains the new $O(\hbar^1)$ renormalised couplings. The new operators take the form
\be 
\cO_{1\,i} \sim \kappa^n h^np^4\,,
\ee
\ie with an extra  $\kappa^2p^2$ compared to $O(\hbar^0)$ vertices. At this stage the arbitrary RG scale $\mu$ is needed so that $\mu^{-2\eps}$ in \eqref{oneloop} can restore dimensions. Since the bare action \eqref{bare} is independent of $\mu$, the renormalised couplings $c_1^{i}$ run with $\mu$. By differentiating \eqref{oneloop} we see that they satisfy:
\be 
\label{beta1}
\beta^i_1 = \dot{c}_1^{i} -2\eps \,c_1^{i} = 2\,a^i_{1/1}\,,
\ee
where $\dot{c}:=\mu\partial_\mu c$.
The one-loop counterterm diagrams formed by using one $a^i_{1/1}$ vertex (corresponding to one copy of $S_1$ being inserted) give in particular double pole divergences 
\be 
\label{counterloop}
\Gamma_{1/2}[S_1] \sim \frac1{\eps^2}\,a_{1/1}\,\mu^{-2\eps}\kappa^{n+2}H^np^{6-2\eps}\,,
\ee
that must satisfy relation \eqref{RGdoublepole}: $\Gamma_{1/2}[S_1]=-2\Gamma_{2/2}$.
As noted by Chase \cite{Chase:1982sf}, the easiest way to see why this is so, is to recognise that the latter take the form
\be 
\label{twoloopdiags}
\Gamma_{2/2}\sim \kappa^{n+2}H^np^{6-4\eps}\left[\frac1{\eps^2}+O\left(\frac1\eps\right)\right]\,,
\ee
but divergences must be local and thus the $(\ln p)/\eps$ terms must cancel between \eqref{counterloop} and \eqref{twoloopdiags}.

We get the same conclusion another way by following Buchler and Colangelo \cite{Buchler:2003vw} whilst also deriving some more useful identities. At $O(\hbar^2)$ the divergences generate the operators
\be 
\cO_{2\,i} \sim \kappa^{n+2} h^np^6\,,
\ee
so we have to add to the bare action
\be 
\label{twoloop}
S_2=\mu^{-4\eps}\int_x\left\{c^i_2\,\cO_{2\,i} +\frac1{\eps^2}a^i_{2/2}\,\cO_{2\,i}+\frac1{\eps}\left(a^i_{2/1}+a^i_{1/1\,j}c^j_1\right)\,\cO_{2\,i}\right\}\,,
\ee
where we now have counterterms with both single and double $\eps$-poles, and $c^i_2$ are the new $O(\hbar^2)$ renormalised couplings. The $a^i_{2/2}$ counterterms cancel the full set of $1/\eps^2$ divergences at $O(\hbar^2)$, \ie from the sum of two-loop diagrams and the one-loop counterterm diagrams.
The single poles $a^i_{2/1}/\eps$ arise from two-loop diagrams using only vertices  \eqref{classical}, whilst the $a^i_{1/1\,j}c^j_1/\eps$ are generated by one-loop diagrams containing one $c_1$ vertex. Now $\mu$-independence of the bare action implies 
\beal 
\beta^i_2 &= \dot{c}^i_2 -4\eps c^i_2 = \frac4\eps a^i_{2/2}+4 \left(a^i_{2/1}+a^i_{1/1\,j}c^j_1\right) -\frac1{\eps} a^i_{1/1\,j}\,\dot{c}^j_1
\,,\nn\\
&= \frac4\eps a^i_{2/2}-\frac2{\eps}a^i_{1/1\,j}a^j_{1/1}+4 a^i_{2/1}+ 2 a^i_{1/1\,j}c^j_1\,,
\label{beta2}
\eeal
where in the second line we substituted the one-loop $\beta$ function \eqref{beta1}. Since this equation is expressed in terms of renormalised quantities, it must be finite, and therefore the single poles must cancel. Thus we see that
\be 
\label{leadingpolerelation}
a^i_{2/2} = \frac12 a^i_{1/1\,j}a^j_{1/1}\,.
\ee
This is the same conclusion as before, but we are now proving it in the form given in \eqref{countertermkey}. The left hand side is the coefficient of the $\cO_{2\,i}$ in $S_{2/2}$ while on the right hand side we have replaced the $c_1^{ j}$ coupling in \eqref{twoloop} by the counterterm coefficient $a^j_{1/1}$. The right hand side is thus the coefficient of $\cO_{2\,i}$ in $-\frac12\Gamma_{1/2}[S_1]$.


Finally let us show that \eqref{countertermkey} will continue to hold when the background, ghosts and antifields are included. Firstly, vertices can now include ghost antighost pairs, but at this schematic level it is not necessary to track these separately from $h$: what really matters in this analysis are the powers of $p^\eps$ and $\mu^\eps$, and they are unchanged if $c$ and $\bar{c}$ are included. Secondly, it is clear that any instance of $h$ (or $H$) can trivially be exchanged for the background $\bar{h}$ in the above schematic formulae, though of course operators $\cO_{1\,j}$ with less than two quantum fields in gauge fixed basis, cannot contribute to the relation \eqref{countertermkey} (their coefficients $a^i_{1/1\,j}$ vanish). Finally from the minimal classical action \eqref{minimal classical action}, we see that whenever an antifield is involved in an action vertex there is one less power of $p$ (compensated dimensionally by the fact that they have $[\phi^*]=2-\eps$, \cf table \ref{table:ghostantighost}). This observation is useful for finding the general form of the corrections, but again for this analysis what actually matters is the tracking of non-integer powers.

\subsection{Properties of the total classical BRST charge}
\label{sec:Properties of the total classical BRST charge}

We now develop the properties of the total classical BRST charge $s_0$. 
Using the identity \cite{Batalin:1981jr,Batalin:1984jr,Batalin:1984ss,Gomis:1994he}:
\be \label{antibracketid} (X,(Y,Z)) = ((X,Y),Z)+(-1)^{(X+1)(Y+1)}(Y,(X,Z))\,,\ee
where $(-1)^X=\pm1$ if $X$ bosonic (fermionic),
we have
\be 
s^2_0X[\phi,\phi^*]=(S_0,(S_0,X))= \half ((S_0,S_0),X) = 0\,,
\ee
where the last equality follows by the CME. Therefore $s_0$ is nilpotent just like the BRST charge $Q$. 
From \eqref{QPhi}, we see that on $\phi^A$ it reduces to the BRST charge $Q$. However from \eqref{s}, $s_0$ also acts on antifields:
\be s_0\phi^*_A=\big(S_0,\phi^*_A\big)=\frac{\partial_r S_0}{\partial\phi^A} \,.\ee
This is called the Koszul-Tate differential \cite{koszul1950type,borel1953cohomologie,tate1957homology,Fisch:1989rp,Morris:2018axr,Kellett:2020mle}. In minimal basis we get explicitly: 
\begin{equation}
    s_0h^{*\mu\nu}=-2\sqrt{g}G^{\mu\nu}/\kappa+2\kappa h^{*\alpha(\mu}\Bar{\nabla}_\alpha c^{\nu)}+\kappa\Bar{\nabla}_\alpha\big(c^\alpha h^{*\mu\nu}\big) \label{sH*}\,,
\end{equation}
\begin{equation}
    s_0c^*_\mu=\kappa\Bar{\nabla}_\mu c^\nu c^*_\nu+\kappa\Bar{\nabla}_\nu\big(c^\nu c^*_\mu\big)-2\Bar{\nabla}_\nu h{^{*\nu}{}_\mu}- 2\kappa \Bar{\nabla}_\alpha\big(h_{\mu\nu}h^{*\alpha\nu}\big)+\kappa\Bar{\nabla}_\mu h_{\alpha\beta}h^{*\alpha\beta} \,.\label{sc*}
\end{equation}
Here $G_{\mu\nu}=-R_{\mu\nu}+\half g_{\mu\nu}R$ is the Einstein tensor. (Note that it inherits an overall minus sign from the Euclidean action compared to the usual definition.) Its indices are raised in \eqref{sH*} using  $G^{\mu\nu}=g^{\mu\alpha}g^{\nu\beta}G_{\alpha\beta}$. As we noted earlier we are raising and lowering indices with the background metric unless explicitly stated otherwise. This case is the one exception.

\begin{table}[ht]
\begin{center}
\begin{tabular}{|c|c|c|c|c|c|c|}
\hline
 & $\epsilon$ & gh \# & ag \# & pure gh \#  & dimension \\
\hline\hline
$h_{\mu\nu}$ & 0 & 0 & 0 & 0 & $(d-2)/2$ \\
\hline
$c_\mu$ & 1 & 1 & 0 & 1 & $(d-2)/2$ \\
\hline \hline
$\bar{c}_\mu$ & 1 & -1 & 1 & 0 & $(d-2)/2$  \\
\hline
$b_\mu$ & 0 & 0 & 1 & 1 & $d/2$\\
\hline\hline\hline
$h^*_{\mu\nu}$ & 1 & -1 & 1 & 0 & $d/2$ \\
\hline
$c^*_\mu$ & 0 & -2 & 2 & 0 & $d/2$\\
\hline\hline
$\bar{c}^*_\mu$ & 0 & 0 & 0 & 0 & $d/2$\\
\hline\hline
$Q$ & 1 & 1 & 0 & 1 & 1 \\
\hline
$Q^-$ & 1 & 1 & -1 & 0 & 1  \\
\hline
\end{tabular}
\end{center}
\caption{The various Abelian charges (\aka gradings) carried by the fields and operators. $\epsilon$ is the Grassmann grading, being $1(0)$ if the object is fermionic (bosonic). gh \# is the ghost number, ag \# the antighost/antifield number, pure gh \# = gh \# + ag \#, and dimension is the engineering dimension. The first two rows are the minimal set of fields, the next two make it up to the non-minimal set, then the ensuing two rows are the minimal set of antifields, and $\bar{c}^*_\mu$ is needed for the non-minimal set. Finally, the charges are determined in order to ensure that  $Q$ and $Q^-$ can also be assigned definite charges.}
\label{table:ghostantighost}
\end{table}

It is useful to assign antighost/antifield number to each field and operator \cite{Morris:2018axr,Fisch:1989rp,Igarashi:2019gkm}, see table \ref{table:ghostantighost}. The reason this is useful is precisely because it is \emph{not} preserved by interactions, which then split into pieces according to their antighost level. For example one sees from \eqref{minimal classical action}, that the three parts of the minimal classical action split into levels 0, 1, and 2, respectively.  The Koszul-Tate differential also splits, in this case into two pieces, one that preserves antighost number and one that lowers it by one. We call these pieces respectively, $Q$ and $Q^-$, and thus write:
\begin{equation}
    s_0\phi^*_A=\big(Q+Q^-\big)\phi^*_A \,.
\end{equation}
From \eqref{sH*} and \eqref{sc*} we see that
\begin{equation}\label{Qh*}
    Qh^{*\mu\nu}=2\kappa h^{*\alpha(\mu}\Bar{\nabla}_\alpha c^{\nu)}+\kappa\Bar{\nabla}_\alpha\big(c^\alpha h^{*\mu\nu}\big)\,,
\end{equation}
\begin{equation}
    Q^-h^{*\mu\nu}=-2\sqrt{g}G^{\mu\nu}/\kappa \label{Q-H*}\,,
\end{equation}
\begin{equation}\label{Qc*}
    Qc^*_\mu=\kappa\Bar{\nabla}_\mu c^\nu c^*_\nu+\kappa\Bar{\nabla}_\nu\big(c^\nu c^*_\mu\big)\,,
\end{equation}
\begin{equation}\label{Q-c*}
    Q^-c^*_\mu=-2\Bar{\nabla}_\nu h{^{*\nu}{}_\mu}- 2\kappa \Bar{\nabla}_\alpha\big(h_{\mu\nu}h^{*\alpha\nu}\big)+\kappa\Bar{\nabla}_\mu h_{\alpha\beta}h^{*\alpha\beta}\,.
\end{equation}
Since $Q$ here acts on antifields there is no reason to confuse it with the previously defined BRST charge \eqref{QPhi}, \eqref{Qbackg}. Its extension to antifields is natural since $Qh^{*\mu\nu}$ and $Qc^*_\mu$ are in fact the correct Lie derivative expressions for these tensor densities. The advantage of the antighost grading becomes clear when we consider the nilpotency of $s_0$:
\begin{equation}
    0=s^2_0=Q^2+\{Q,Q^-\}+(Q^-)^2\,.
\end{equation}
These terms must vanish separately since they lower the antighost number by $0$, $1$ and $2$ respectively. Therefore we know that our definitions of $Q$ and $Q^-$ are such that they are nilpotent and they anticommute. 

\subsection{Canonical transformations up to second order}
\label{sec:canonsecond}

We saw in sec. \ref{sec:RGandCME} that a central r\^ole is played by counterterms that are $s_0$-closed, for example at one loop we have exactly this relation \eqref{SoneCME}: $s_0S_1=0$. We saw in the previous subsection that $s_0$ is nilpotent, so one solution to this is that $S_1$ is exact: $S_1=s_0K_1$, where $K_1$ is a local functional of ghost number $-1$. In the next subsection we derive the general solution for such $s_0$-closed counterterms, but for that we will need the relation between $s_0$-exact solutions and canonical transformations. Taking the general canonical transformation \eqref{canon}, and setting 
\be \K = \check{\phi}^*_A \phi^A +K_1[\phi,\check{\phi}^*]\,,\ee 
and then treating $K_1$ to first order, one gets the following field and source reparametrisations
\begin{equation}\label{canonsmall}
    \delta \phi^A=\frac{\partial_lK_1}{\partial \phi^*_A}\,, \qquad \delta \phi^*_A=-\frac{\partial_lK_1}{\partial \phi^A}\,.
\end{equation}
That these correspond to $s_0$-exact solutions, can then be seen by writing out the change in the classical action:
\begin{equation}
    \delta S_0=\frac{\partial_r S_0}{\partial\phi^A}\delta \phi^A+\frac{\partial_r S_0}{\partial \phi^*_A}\delta\phi^*_A=\frac{\partial_r S_0}{\partial\phi^A}\frac{\partial_lK_1}{\partial \phi^*_A}-\frac{\partial_r S_0}{\partial \phi^*_A} \frac{\partial_lK_1}{\partial \phi^A}=s_0K_1\,.
\end{equation}

This interpretation extends to higher orders \cite{Gomis:1995jp}, see also \cite{Dixon:1975si,Voronov:1982cp,Voronov:1982ph,Lavrov:1985ugv,Anselmi:1994zx}. For sec. \ref{sec:solodukhin} we will want their explicit form to second order. Given that $S_1=s_0K_1$, one solution to the CME to second order \eqref{StwoCME}, \ie $s_0S_2=-\half(S_1,S_1)$, is:
\be S_2 = \frac12(S_1,K_1)+s_0K_2 \label{Stwogensol}\ee
where $K_2$ is a second-order local functional of ghost number -1. This follows from the antibracket identity \eqref{antibracketid} because 
\be s_0(S_1,K_1) =(s_0S_1,K_1)-(S_1,s_0K_1) = -(S_1,S_1)\,.\label{solvingabracket}\ee
In fact the relation \eqref{Stwogensol} is just the result of taking the $K_1$ canonical transformation to second order and adding the new part $K_2$ which appears linearly at this order. To see this we set
\be \K = \check{\phi}^*_A \phi^A +K_1[\phi,\check{\phi}^*]+K_2[\phi,\check{\phi}^*]\,,\ee
and solve the exact canonical transformation \eqref{canon} perturbatively for $\delta\phi^{(*)}=\check{\phi}^{(*)}-\phi^{(*)}$, starting with the first order expression \eqref{canonsmall}. We get
\beal 
\delta \phi^A &=\frac{\partial_lK_1}{\partial \phi^*_A} +\frac12\frac{\partial_l}{\partial\phi^*_A}\frac{\partial_rK_1}{\partial\phi^B}\frac{\partial_lK_1}{\partial\phi^*_B}-\frac12\frac{\partial_l}{\partial\phi^*_A}\frac{\partial_rK_1}{\partial\phi^*_B}\frac{\partial_lK_1}{\partial\phi^B}+\frac{\partial_lK_2}{\partial\phi^*_A}
\,,\nn\\
\quad \delta \phi^*_A &=-\frac{\partial_lK_1}{\partial \phi^A} +\frac12\frac{\partial_l}{\partial\phi^A}\frac{\partial_rK_1}{\partial\phi^*_B}\frac{\partial_lK_1}{\partial\phi^B}-\frac12\frac{\partial_l}{\partial\phi^A}\frac{\partial_rK_1}{\partial\phi^B}\frac{\partial_lK_1}{\partial\phi^*_B}-\frac{\partial_lK_2}{\partial\phi^A}\,.\label{canonsecond}
\eeal
Taylor expanding the classical action to second order gives
\beal \delta S_0\ =\ &\frac{\partial_rS_0}{\partial\phi^A}\delta\phi^A +\frac12\frac{\partial_r}{\partial\phi^B}\left(\frac{\partial_rS_0}{\partial\phi^A}\delta\phi^A\right)\delta\phi^B+\frac12\frac{\partial_r}{\partial\phi^*_B}\left(\frac{\partial_rS_0}{\partial\phi^A}\delta\phi^A\right)\delta\phi^*_B\nn\\
+&\frac{\partial_rS_0}{\partial\phi^*_A}\delta\phi^*_A +\frac12\frac{\partial_r}{\partial\phi^B}\left(\frac{\partial_rS_0}{\partial\phi^*_A}\delta\phi^*_A\right)\delta\phi^B+\frac12\frac{\partial_r}{\partial\phi^*_B}\left(\frac{\partial_rS_0}{\partial\phi^*_A}\delta\phi^*_A\right)\delta\phi^*_B\nn\\
&\phantom{\frac{\partial_rS_0}{\partial\phi^A}\delta\phi^A}-\frac12\frac{\partial_rS_0}{\partial\phi^A}\left(\frac{\partial_r}{\partial\phi^B}\delta\phi^A\right)\delta\phi^B-\frac12\frac{\partial_rS_0}{\partial\phi^A}\left(\frac{\partial_r}{\partial\phi^*_B}\delta\phi^A\right)\delta\phi^*_B\nn\\
&\phantom{\frac{\partial_rS_0}{\partial\phi^*_A}\delta\phi^*_A}-\frac12\frac{\partial_rS_0}{\partial\phi^*_A}\left(\frac{\partial_r}{\partial\phi^B}\delta\phi^*_A\right)\delta\phi^B-\frac12\frac{\partial_rS_0}{\partial\phi^*_A}\left(\frac{\partial_r}{\partial\phi^*_B}\delta\phi^*_A\right)\delta\phi^*_B\,.
\eeal
Substituting \eqref{canonsecond}, its non-linear terms cancel the final two lines, whilst the first two lines organise into antibrackets, and thus we find that 
\be \delta S_0 = (S_0,K_1+K_2) +\frac12((S_0,K_1),K_1) = s_0K_1+\frac12(S_1,K_1)+s_0K_2\,,\label{STaylor}\ee
showing that the non-linear term in \eqref{Stwogensol}, is indeed the result \eqref{canonsecond} of carrying the canonical transformation to second order.

\subsection{General form of $s_0$-closed divergences} 
\label{sec:General form of divergences in the Legendre effective action}

On the other hand, at each new loop order the $s_0$-closed counterterms are associated to the `new' part $\Gamma_\infty$ of the divergences. Their form can be classified by the cohomology of $s_0$ in the space of local functionals. As we have seen, one possibility is that it is a local $s_0$-exact solution: $\Gamma_\infty=s_0K_\infty[\Phi,\Phi^*]$, where $K_\infty$ is a functional with ghost number $-1$. However another possibility is that the divergence is a local functional $\Gamma_\infty[g_{\mu\nu}]$ of only the total metric,\footnote{\label{foot:caps} We write the vacuum expectation value of the quantum fields in capitals, thus in minimal basis $\Phi^A=H_{\mu\nu},C^\xi$.} $g_{\mu\nu}=\bar{g}_{\mu\nu}+\kappa H_{\mu\nu}$, and is diffeomorphism invariant.
Ref. \cite{Barnich:1994kj}, see also \cite{Gomis:1995jp}, proves from the cohomological properties of $s_0$ that if the background metric is flat, \viz $\bar{g}_{\mu\nu}=\delta_{\mu\nu}$, then in fact the general local $s_0$-closed solution is a linear combination of these two possibilities:
\be 
\label{gensoln}
s_0\Gamma_\infty[\Phi,\Phi^*] = 0 \qquad \implies \qquad \Gamma_\infty[\Phi,\Phi^*] = \Gamma_\infty[g_{\mu\nu}]+s_0K_\infty[\Phi,\Phi^*]\,.
\ee
However in a non-flat background, as a statement on $s_0$-cohomology, this result is no longer true, since clearly one can now add to this a local functional $\Gamma_\infty[\bar{g}_{\mu\nu}]$ of only the background field (such a functional being trivially annihilated by $s_0$). 
Nevertheless it is true as a statement about $s_0$-closed divergences, as we show below.

Before doing so, we note that it is useful in this paper to grade the solution \eqref{gensoln} by antighost number. The first part, $\Gamma[g]$, has of course zero antighost number, but since $K$ has ghost number $-1$, we see from table \ref{table:ghostantighost} that it splits up as $K=K^1+K^2+\cdots$, where the superscript denotes antighost number. Thanks to the perturbative non-renormalisability of quantum gravity, already at one loop one finds that all these infinitely many $K^n$ functionals are non-vanishing. In minimal basis, $K^1$ is characterised by having one copy of $H^*$, $K^2$ by containing one copy of $C^*$ or two copies of $H^{*}$ whilst also being linear in the ghost $C^\mu$, and so on, with the higher level $K^n$ containing ever greater numbers of antifields and compensating powers of ghosts.

Now we show that \eqref{gensoln} is indeed the general form of an $s_0$-closed divergence, even in a non-trivial background. Although this is effectively a small extension of the proof in flat background, it has not, to our knowledge, been noticed before.
Following \cite{Abbott:1980hw}, first we observe that, up to a choice of gauge, the Legendre effective action can equivalently be computed by shifting 
\be \label{shifted} h_{\mu\nu}\mapsto h_{\mu\nu}-\bar{h}_{\mu\nu}\ee  
which, by \eqref{gback expansion}, amounts to
expanding around flat space. Indeed this shift makes no difference to the minimal classical action \eqref{minimal classical action}, since it depends only on the total metric $g_{\mu\nu}$. Differences arise only because separate $h_{\mu\nu}$ and $\bar{g}_{\mu\nu}$ dependence enters via the canonical transformation induced by the gauge fixing fermion  \eqref{BF gauge fixing fermion}, which from (\ref{canon},\ref{canonpsi}) takes the form
\be Q\phi^A\frac{\partial\Psi}{\partial\phi^A} = Q\Psi[\phi]\,,\label{gauge}\ee
and enters via the quadratic $b^\mu$ term from the extension \eqref{extended action}, which can however also be written in $Q$-exact form
\be\label{bchange} \frac{1}{2 \alpha}\sqrt{\Bar{g}} \Bar{g}_{\mu\nu} b^\mu b^\nu = \frac{i}{2 \alpha} Q\left(\sqrt{\Bar{g}} \Bar{g}_{\mu\nu} \bar{c}^\mu b^\nu\right) = Q\Psi_b[\phi]\,. \ee
Thus the entire $\bar{g}$ (equivalently $\bar{h}$) dependence can be seen as being just part of the parametrisation of our choice of gauge, \ie of $\Psi_\text{tot}[\phi]=\Psi[\phi]+\Psi_b[\phi]$. 

Now in the shifted basis \eqref{shifted} we are expanding around flat space. If we also use  an $\bar{h}$-independent gauge, then we can be sure that \eqref{gensoln} holds. We cannot use this result directly  to rule out a separate $\Gamma_\infty[\bar{g}_{\mu\nu}]$ piece, because we have changed the gauge. However we can proceed by comparing physical quantities since they are independent of the choice of gauge. We do this by setting $\Phi^*_A=0$ and setting $H_{\mu\nu}$ on shell. Note that since we are dealing with new divergences appearing at some given loop order, it is the classical equations of motion for $g_{\mu\nu}$ that one needs. Then $\Gamma_\infty[g_{\mu\nu}]$ is independent of the background, whilst $s_0K_\infty$ vanishes. The latter follows because
\be s_0K_\infty = \frac{\partial_r S_0}{\partial \Phi^A} \frac{\partial_l K_\infty}{\partial \Phi^*_A} - \frac{\partial_r S_0}{\partial \Phi^*_A} \frac{\partial_l K_\infty}{\partial \Phi^A} \,.\ee
Given that $\Phi^*_A=0$, on the right hand side the first term vanishes (in minimal basis) by the equations of motion of $H_{\mu\nu}$, and the second term because $K_\infty$ has non-vanishing antighost number. Now comparing the results in flat background and non-flat background, we see that they must have the same total metric part $\Gamma_\infty[g_{\mu\nu}]$, whilst for a non-flat background the purely background part must vanish:  $\Gamma_\infty[\bar{g}_{\mu\nu}]=0$.

We finish with some important remarks. Firstly, to avoid over-counting, the counterterm $S_\ell[g]$ for the pure metric part of the $s_0$-closed solution \eqref{gensoln} should be restricted to terms that do not vanish on the classical equations of motion (or more generally to a specific choice, as in \eqref{GB}, the Gauss-Bonnet term). To see this we note that if $S_\ell[g]$ does vanish on the classical equations of motion, it can be written as
\begin{equation}
    S_\ell[g_{\mu\nu}]=-\frac{2}{\kappa}\int_x \sqrt{g}\,G^{\mu\nu}T_{\mu\nu}[g_{\mu\nu}]=Q^-\int_x h^{*\mu\nu}T_{\mu\nu}=s_0\int_x h^{*\mu\nu}T_{\mu\nu} \label{exact}
\end{equation}
for some tensor $T_{\mu\nu}[g_{\mu\nu}]$. In the last step we used the fact that both $h^{*\mu\nu}$ and $T_{\mu\nu}$ transform properly as tensor densities under $Q$. Thus any part of $S_\ell[g]$ that vanishes on the classical equations of motion can be written instead as part of the $s_0$-exact piece, $s_0K_\ell$, \ie to a canonical transformation taken to first order.

Secondly, notice that it is important for the above arguments that we are setting $H_{\mu\nu}$ on shell, but not the background metric $\bar{g}_{\mu\nu}$. This is what allows us to deduce that there cannot be any purely background part. On the other hand in the background field method one sets all the classical fields to zero and keeps only the background metric. Although this technique is not the primary focus of the paper (apart from in sec. \ref{sec:solodukhin}) the proof here tells us something important about it. Since on shell, the background field effective action gives the same results  \cite{Abbott:1980hw}, we know that divergences that do not vanish on the background equations of motion descend from functionals of the total metric $g_{\mu\nu}$, whilst those divergences that vanish on the background equations of motion belong to canonical transformations and are thus removed by reparametrising $h_{\mu\nu}$ not the background field.

\section{Explicit expressions for counterterms} 
\label{sec:Loops}

We now verify these results in explicit loop computations, up to two loops, in particular we draw out the intimate relationship between the leading off-shell divergences for the two-point vertex up to two loops and the one-loop counterterm diagrams \cite{Buchler:2003vw} and in turn to canonical transformations in the BRST algebra \cite{Gomis:1995jp}. Since this necessitates computing, as an intermediate step, the off-shell divergences in one-loop diagrams with three external legs, two of which are fluctuation fields, we widened our investigation so as to compute explicitly all off-shell one-loop divergences with up to three (anti)fields. Below we express these divergences in terms of the minimal-basis counterterms in $S_\ell$ ($\ell=1,2$) that one needs to add to the bare action. In minimal subtraction, which we follow, the counterterms are just minus the divergences. However, since the bare action is a $\mu$-independent local functional, the RG and CME relationships are most naturally expressed in terms of the counterterms, as we have seen in secs. \ref{sec:RGandCME} and \ref{sec:RG}.

In fact it was in the process of computing these that we noticed that purely background metric pieces were not generated, which motivated the general proof in sec. \ref{sec:General form of divergences in the Legendre effective action}. It was also whilst analysing these that we noticed that the RG relations for counterterms are actually crucial for consistency of the BRST algebra as realised on the Legendre effective action versus as realised on the counterterms. This is explained in sec. \ref{sec:RGandCME}. Finally these results allowed explicit verification that the generalised $\beta$ function proposal of ref. \cite{Solodukhin:2020vuw} cannot be correct, which led to us formulating the detailed analysis provided in sec. \ref{sec:solodukhin}. We similarly hope that these examples will prove useful in future studies of perturbative quantum gravity.

Just like for $K$ in sec. \ref{sec:General form of divergences in the Legendre effective action}, it is useful to split the Legendre effective action and bare action according to antighost number. All antighost levels $S^n$ depend on the graviton fields $h_{\mu\nu}$ and $\bar{h}_{\mu\nu}$, but their dependence on (anti)ghosts is restricted by the quantum numbers, \cf table \ref{table:ghostantighost}. Thus $S^0$ depends only on the graviton fields, whilst $\Gamma^0[H_{\mu\nu},\bar{H}_{\mu\nu}]$ is the physical part that ultimately provides the S-matrix, $S^1$ is linear in $h^{*\mu\nu}$ and $c^\mu$ (in gauge fixed basis \eqref{gf to gi 1}, $S^1$ renormalises the ghost action), $S^2$ is made of vertices containing two $c^\mu$ and either one $c^*_\mu$ or two $h^{*\alpha\beta}$, and so on.

\subsection{One-loop two-point counterterms}
\label{sec:onelooptwopt}

\subsubsection{Level zero, \ie graviton, counterterms}
\label{sec:onelooptwoptagzero}
Recall from sec. \ref{sec:The CME for the bare action} that we are using Feynman DeDonder gauge.
As explained in the next subsection, in this case it turns out that the result for the one-loop two-point graviton counterterm can be expressed entirely in terms of curvatures linearised around the flat metric. In particular let us introduce for the quantum fluctuation the linearised `quantum curvature'
\be \label{deflinearisedcurvature} R_{\mu\alpha\nu\beta} =\kappa \R_{\mu\alpha\nu\beta} +O(\kappa^2)\,,\ee
where we are expanding $g_{\mu\nu}=\delta_{\mu\nu}+\kappa h_{\mu\nu}$,
and thus 
\be
\label{curvatures}
R^{(1)}_{\mu\alpha\nu\beta} = -2 \partial_{[\mu|\,}\partial_{[\nu} h_{\beta]\,|\alpha]}\,,\  R^{(1)}_{\mu\nu} = -\partial^2_{\mu\nu}\vp+\partial_{(\mu}\partial^\alpha h_{\nu) \alpha}-\half\, \Box\, h_{\mu\nu}\,,\  R^{(1)} = \partial^2_{\alpha\beta}h_{\alpha\beta}-2\,\Box\,\vp
\ee
(defining $\half (t_{\mu\nu}\pm t_{\nu\mu})$ for symmetrisation $t_{(\mu\nu)}$, respectively antisymmetrisation $t_{[\mu\nu]}$).
Here we are using $\ph=\half\delta^{\mu\nu}h_{\mu\nu}$\footnote{This definition is the previous one \eqref{phi} after linearisation.} and indices are raised and lowered with the flat metric $\delta_{\mu\nu}$. Following the definition below \eqref{sc*}, the linearised Einstein tensor is then $G^{(1)}_{\mu\nu}=-\R_{\mu\nu}+\half \delta_{\mu\nu}\R$. 
Similarly we introduce the corresponding linearised background curvatures $\bR_{\mu\alpha\nu\beta}$ \etc and linearised background Einstein tensor $\bar{G}^{(1)}_{\mu\nu}$, by  replacing $h_{\mu\nu}$ with $\bar{h}_{\mu\nu}$. 

\ignore{
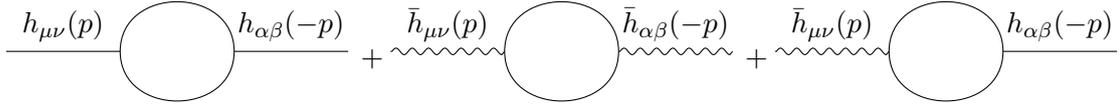
\begin{figure}[ht]
\centering
$$
\begin{gathered}
\begin{tikzpicture}
  \begin{feynman}
    \vertex (a) ;
    \vertex [right=of a] (b);
     \vertex [right=of b] (c);
      \vertex [right=of c] (d);
\diagram* {
      (a) -- [plain, edge label=$h_{\mu\nu}(p)$] (b), (b) -- [plain, half right] (c),
      (c) -- [plain, edge label=$h_{\alpha\beta}(-p)$] (d), (b)
-- [plain, half left] (c),
    }; 
  \end{feynman}
\end{tikzpicture}
\end{gathered}
+
\begin{gathered}
\begin{tikzpicture}
  \begin{feynman}
    \vertex (a) ;
    \vertex [right=of a] (b);
     \vertex [right=of b] (c);
      \vertex [right=of c] (d);
\diagram* {
      (a) -- [photon, edge label=$\Bar{h}_{\mu\nu}(p)$] (b), (b) -- [plain, half right] (c),
      (c) -- [photon, edge label=$\Bar{h}_{\alpha\beta}(-p)$] (d), (b)
-- [plain, half left] (c),
    }; 
  \end{feynman}
\end{tikzpicture}
\end{gathered}
+
\begin{gathered}
\begin{tikzpicture}
  \begin{feynman}
    \vertex (a) ;
    \vertex [right=of a] (b);
     \vertex [right=of b] (c);
      \vertex [right=of c] (d);
\diagram* {
      (a) -- [photon, edge label=$\Bar{h}_{\mu\nu}(p)$] (b), (b) -- [plain, half right] (c),
      (c) -- [plain, edge label=$h_{\alpha\beta}(-p)$] (d), (b)
-- [plain, half left] (c),
    }; 
  \end{feynman}
\end{tikzpicture}
\end{gathered}
$$
\caption{Two-point graviton diagrams at one loop. The wavy line represents the background field and the external plain line represents the quantum graviton field. The internal lines represent both a graviton loop and a ghost loop.}
\label{fig:twopoints}
\end{figure}}
Computing the diagrams in fig. \ref{fig:twopoints} we find
\besp
\label{oneloopzerotwopt}
S^0_{1/1} = \frac{\kappa^2\mu^{-2\eps}}{(4\pi)^2\eps}\int_x \Big\{
\frac{61}{60}(R^{(1)}_{\mu\nu})^2-\frac{19}{120}(R^{(1)})^2 
+\frac{7}{20} (\Bar{R}^{(1)}_{\mu\nu})^2+\frac{1}{120}(\Bar{R}^{(1)})^2 \\
+\frac{41}{30}R^{(1)}_{\mu\nu}\Bar{R}^{(1)\mu\nu} - \frac{3}{20}R^{(1)}\Bar{R}^{(1)}
\Big\}\,.
\eesp
The first diagram gives the first two terms, \ie the pure quantum terms. The result agrees with ref. \cite{Kellett:2020mle}. It was calculated in a general two parameter gauge in ref. \cite{Capper:1979ej}. After correcting some typos and specialising to Feynman DeDonder gauge, it also agrees. The next two terms, the purely background terms, agree with the famous result in \cite{tHooft:1974toh} and (up to a factor of $1/2$) with \cite{Goroff:1985sz}. For more details on these comparisons, see app. \ref{app:comparisons}. To our knowledge the last two terms, \ie the mixed terms, have not appeared in the literature before. 

By \eqref{SoneCME}, the terms \eqref{oneloopzerotwopt} must be part of an $s_0$-closed counterterm action $S_{1/1}$. Furthermore 
according to the proof given in sec. \ref{sec:General form of divergences in the Legendre effective action}, since the quantum curvature pieces vanish on the equations of motion and since there cannot be a separate purely background part, we must be able to express the entire result as $s_0$-exact, and thus in fact the terms must collect into 
\be \label{exactoneloopzerotwopt} S^0_{1/1} = Q^- K^1_{1/1}\,.\ee 
Given that \eqref{oneloopzerotwopt} is made solely of linearised curvatures, at the two-point level the only possible terms in $K^1_{1/1}$ that can contribute, are:
\be \label{K1ag1-2pt} K^1_{1/1}\ \ni\ 
\frac{\kappa^2\mu^{-2\eps}}{(4\pi)^2\eps}\int_x \left\{ \beta h^{*\mu\nu}R^{(1)}_{\mu\nu}+\gamma\varphi^*R^{(1)}
+\Bar{\beta} h^{*\mu\nu}\Bar{R}^{(1)}_{\mu\nu}+\Bar{\gamma}\varphi^*\Bar{R}^{(1)}\right\}\,,\ee
where $\beta$, $\gamma$, $\bar{\beta}$ and $\bar{\gamma}$ are parameters to be determined, and we have introduced 
\be \ph^* = \half \bar{g}_{\mu\nu}h^{*\mu\nu}\ee 
by analogy with \eqref{phi} (although here $\bar{g}_{\mu\nu}$ can be replaced by $\delta_{\mu\nu}$). It is apparent that we have six numbers in \eqref{oneloopzerotwopt} to reproduce with only four parameters, and therefore this relation is a non-trivial check on the formalism.  From \eqref{Q-H*}, the action of $Q^-$ reduces in this case to 
\be\label{Q-h*linearised} Q^-h^{*\mu\nu} = -2\left( G^{(1)\mu\nu}+\bar{G}^{(1)\mu\nu}\right)\,,\ee
and thus from \eqref{exactoneloopzerotwopt} and \eqref{K1ag1-2pt},
\besp \label{Sonebetagammas}
    S^0_{1/1}= \frac{\kappa^2\mu^{-2\eps}}{(4\pi)^2\eps}\int_x \Big\{2\beta (R^{(1)}_{\mu\nu})^2-[\beta+\gamma](R^{(1)})^2+2\Bar{\beta}(\Bar{R}^{(1)}_{\mu\nu})^2-[\Bar{\beta}+\Bar{\gamma}](\Bar{R}^{(1)})^2+2[\beta+\Bar{\beta}]R^{(1)}_{\mu\nu}\Bar{R}^{(1)\mu\nu} \\
    -[\beta+\Bar{\beta}+\gamma+\Bar{\gamma}]R^{(1)}\Bar{R}^{(1)}\Big\}\,.
\eesp
We see that the mixed Ricci-squared terms must have a coefficient which is simply the sum of the coefficients of the pure quantum and pure background Ricci-squared terms, and likewise for the  scalar-curvature-squared terms. The reader can verify from \eqref{oneloopzerotwopt} that these two constraints are indeed satisfied. Therefore there are four independent constraints and we can find a consistent (and unique) solution. It is:
\be\label{betagammavals} \beta = \frac{61}{120}\,,\quad \gamma = -\frac{7}{20}\,,\quad \Bar{\beta}=\frac{7}{40}\,, \quad \Bar{\gamma}=-\frac{11}{60}\,.\ee

\subsubsection{Level one (\aka ghost) counterterms}
\label{sec:onelooptwoptagone}

The level one two-point counterterm is computed by using the classical three-point vertices involving $h^{*\mu\nu}$, and transferring to gauge fixed basis using \eqref{gf to gi 1}. We display the result in minimal basis where it takes its simplest form, since it then contains only the divergent corrections to $Qh_{\mu\nu}$ (at the linearised level, compare \eqref{onelooponetwopt} to \eqref{minimal classical action} and \eqref{Qbackg}), but in gauge fixed basis the generated $\bar{c}^\alpha$ terms are the counterterms necessary to renormalise the ghost action, \eqref{g.f. action}. We find that
\be S^1_{1/1} = \frac{\kappa^2\mu^{-2\eps}}{(4\pi)^2\eps}\int_x \left\{ \frac12 h^{*\mu\nu}\partial^3_{\mu\nu\alpha}c^\alpha-\frac34h^{*\mu\nu}\Box\partial_\mu c_\nu \right\}\,,\label{onelooponetwopt} \ee
in agreement with ref. \cite{Kellett:2020mle}, \cf app. \ref{app:comparisons}.
Again these must belong to $s_0K_{1/1}$ for a suitable choice of $K_{1/1}$, which means that we must add to what we have in \eqref{K1ag1-2pt}. A solution is to add
\be K^1_{1/1}\ \ni\ -\frac12\frac{\kappa^2\mu^{-2\eps}}{(4\pi)^2\eps}\int_x h^{*\mu\nu} \partial^2_{\mu\nu}\ph\,,\qquad  K^2_{1/1}\ \ni\ -\frac38\frac{\kappa^2\mu^{-2\eps}}{(4\pi)^2\eps}\int_x c^*_\mu\Box c^\mu\,. \label{K1ag12-2pt}\ee
At the two-point level, it is straightforward to see that the level-two part gives the second term in \eqref{onelooponetwopt} via $Q^-K^2_{1/1}$, whilst the level-one part gives the first term via $QK^1_{1/1}$, \eqref{K1ag1-2pt} making no contribution because it is annihilated by $Q$. On the other hand, \eqref{K1ag1-2pt} is still correct for reproducing $S^0_{1/1}$ because the level-one part above is annihilated by $Q^-$, as follows by the Bianchi identity for the Einstein tensor or by recognising that the above level-one part is proportional to $Q^-(\partial^\alpha c^*_\alpha\ph)$. Indeed at this stage one has to face the issue that the solution for $K$ is unique only in the cohomology. One can always add an $s_0$-exact piece to $K$, in particular one can add $s_0(\partial^\alpha c^*_\alpha\ph)$. The above solution is one choice, in fact the same as that made in ref. \cite{Kellett:2020mle}.

Now let us comment on the results of the previous subsection. The fact that they can be written covariantly, in terms of curvatures of the background metric, is of course no accident: this is guaranteed by background diffeomorphism invariance. The fact that one can also do so in terms of $g_{\mu\nu}=\delta_{\mu\nu}+\kappa h_{\mu\nu}$, is however an accident of Feynman DeDonder gauge. At the level of the action it is a consequence of the fact that $Q^- S^1_{1/1}=0$ in this gauge, and thus the graviton counterterm action must be annihilated by $Q$:
\be 0=s_0 S_{1/1} = Q S^0_{1/1}+Q^- S^1_{1/1}= QS^0_{1/1}\,.\ee  
Up to cohomology and normalisation, there is a unique term  $\ph^*\Box\ph\in K_{1/1}$ that could arise in the one-loop calculation which would break this `quantum diffeomorphism' invariance. Equivalently in $S^0_1$ we would find a term proportional to
\be \ Q^- \int_x \ph^*\Box\ph = -\int_x (\R+\bR)\Box\ph\,.\label{oneloopQbreaking}\ee
Indeed from  \cite{Capper:1979ej}, \cf app. \ref{app:comparisons}, we know this term is present in a more general gauge. Furthermore we will see in sec. \ref{sec:twoloops} that at two loops an analogous term is generated even in Feynman DeDonder gauge, while at one loop but beyond the two-point level many terms ensure that $QS^0_{1/1}\ne0$.

This completes the calculation at the two-point level because it is not possible to generate two-point higher level counterterms $S^{n>1}_\ell$ (since $n$ is also the pure ghost number). 

\subsection{One-loop three-point counterterms}
\label{sec:threepoints}

This involves computing one-loop diagrams with the topologies given in fig. \ref{fig:threepoint}. 
Already at this stage there are thousands of divergent vertices, and computer algebra becomes essential.  We proceed by comparing the results with the general structure \eqref{gensoln}, \ie we should find that the counterterm action takes the form:
\be \label{oneloopgensoln} S_{1/1}[\Phi,\Phi^*] = S^0_{1/1}[g_{\mu\nu}]+s_0K_{1/1}[\Phi,\Phi^*]\,.\ee

\ignore{
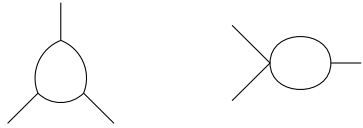
\begin{figure}[ht]
\centering
$$
\begin{gathered}
\begin{tikzpicture}
\begin{feynman}
\vertex  (a) at (0,0);
\vertex (b) at (0,0.4);
\vertex (c) at (-0.3,-0.3);
\vertex (d) at (0.3,-0.3);
\vertex   (b1) at (0,0.9);
\vertex   (c1) at (-0.7,-0.7);
\vertex   (d1) at (0.7,-0.7); 
\diagram*{
(b) -- [plain, quarter right] (c),
(c) -- [plain, quarter right] (d),
(d) -- [plain, quarter right] (b),
(b) -- [plain] (b1),
(c) -- [plain] (c1),
(d) -- [plain] (d1),
};
\end{feynman}
\end{tikzpicture}
\end{gathered}
\qquad\qquad
\begin{gathered}
\begin{tikzpicture}
\begin{feynman}
\vertex  (a) at (0,0);
\vertex (b) at (-0.5,0.5);
\vertex  (c) at (-0.5,-0.5);
\vertex (d) at (0.8,0);
\vertex  (e) at (1.2,0);
\diagram*{
(a) -- [plain, half right] (d),
(d) -- [plain, half right] (a),
(a) -- [plain] (b),
(a) -- [plain] (c),
(d) -- [plain] (e),
};
\end{feynman}
\end{tikzpicture}
\end{gathered}
$$
\caption{Three-point Feynman diagrams at one loop.}
\label{fig:threepoint}
\end{figure}}

As explained in ref. \cite{Goroff:1985th}, dimensional regularisation allows for the computation of the Gauss-Bonnet topological term:
\be\label{GB} S^0_{1/1}[g_{\mu\nu}] = \frac{\tau\mu^{-2\eps}}{(4\pi)^2\eps} \int_x\! \sqrt{g} \left( R^{\mu\nu\rho\sigma}R_{\mu\nu\rho\sigma}+R^2-4R^{\mu\nu}R_{\mu\nu}\right) = \frac{\tau\mu^{-2\eps}}{(8\pi)^2\eps}\int_x \!\sqrt{g}\,\epsilon^{\alpha\beta\gamma\delta}\epsilon_{\mu\nu\rho\sigma}R^{\mu\nu}_{\phantom{\mu\nu}\alpha\beta}R^{\rho\sigma}_{\phantom{\rho\sigma}\gamma\delta}\,,\ee
which is the unique possibility for $S^0_{1/1}[g_{\mu\nu}]$ up to choice of coefficient $\tau$ and terms that vanish on shell (\cf the discussion at the end of sec. \ref{sec:General form of divergences in the Legendre effective action}). 

Up to the three-point level, $K_{1/1}$ has no more than antighost number two. The two 
antighost levels have the following general parametrisation:
\begin{align}
    K^1_{1/1}\ =\ &\frac{\kappa\mu^{-2\eps}}{(4\pi)^2\eps}\int_x \left( \Bar{\beta} h^{*\mu\nu}\Bar{R}_{\mu\nu}+\Bar{\gamma}\varphi^*\Bar{R}\right)
+\frac{\kappa^2\mu^{-2\eps}}{(4\pi)^2\eps}\int_x \Big\{\beta h^{*\mu\nu}\left( \Bar{\nabla}_\mu\Bar{\nabla}^\alpha  h_{\alpha\nu}-\tfrac12\Bar{\square}  h_{\mu\nu}\right)\nn\\
&+\left(c_1-\beta\right)h^{*\mu\nu}\Bar{\nabla}_\mu \Bar{\nabla}_\nu \varphi
+\gamma\varphi^*\left(\Bar{\nabla}_\alpha\! \Bar{\nabla}_\beta  h^{\alpha\beta} 
-2\Bar{\square}\varphi\right)+ \alpha_3  h^{*\mu\nu}\Bar{R}_\mu{}^\alpha  h_{\alpha\nu}+\alpha_4 h^{*\mu\nu}\Bar{R}_{\alpha\mu\nu\beta} h^{\alpha\beta}
\nn\\ &+  \alpha_5  h^{*\mu\nu}\Bar{R}_{\mu\nu}\varphi
    +   \alpha_6\varphi^*\Bar{R}^{\alpha\beta} h_{\alpha\beta}+  \alpha_7 \Bar{R}\varphi^*\varphi\Big\} + \frac{\kappa^3\mu^{-2\eps}}{(4\pi)^2\eps}\int_x\sum_{i=1}^{27} b_i  \left(h^{*}h^2\partial^2\right)_i   \,,\label{K11}
\end{align}
\begin{align}
    K^2_{1/1}\ =\ &\frac{\kappa^2\mu^{-2\eps}}{(4\pi)^2\eps}\int_x \left(c_2c^*_\mu \Bar{\square} c^\mu +\alpha_1 c^*_\mu c^\mu \Bar{R}+\alpha_2 c^*_\mu c^\nu \Bar{R}^\mu{}_\nu\right) +\frac{\kappa^3\mu^{-2\eps}}{(4\pi)^2\eps}\int_x\frac{1}{\sqrt{\bar{g}}}\Big(   \alpha_8 \varphi^*\Bar{\nabla}_\mu \varphi^*c^\mu \nonumber \\ 
    &+  \alpha_9 h^{*\alpha\beta}\Bar{\nabla}_\mu   h^*_{\alpha\beta}c^\mu +   \alpha_{10} \varphi^* h^{*}_{\mu\nu}\Bar{\nabla}^\mu c^\nu +   \alpha_{11}  h^{*}_{\alpha\mu} h^{*\alpha}{}_\nu \Bar{\nabla}^\mu c^\nu \Big) +\frac{\kappa^3\mu^{-2\eps}}{(4\pi)^2\eps}\int_x\sum^{21}_{i=1} d_i \left(c^*ch\partial^2\right)_i \,.\label{K21}
\end{align}
Here we have used the symmetries and statistics of the (anti)fields. In particular, the result must be background diffeomorphism invariant (which implies the factor of $1/\sqrt{\bar{g}}$ in the terms with two antifields, because we defined them to transform as tensor densities of weight $-1$). Furthermore, we know that the terms with one antifield have two space-time derivatives whilst those with two antifields have one spacetime derivative. The power of $\kappa$ and $\mu$ then follow from $[K]=-1$. 

The parametrisation must be consistent with the results at the two-point level, hence the appearance of parameters $\bar{\beta}$, $\bar{\gamma}$, $\beta$ and $\gamma$ from \eqref{K1ag1-2pt}. We similarly introduce parameters $c_1$ and $c_2$ where, from \eqref{K1ag12-2pt}, we know that 
\be \label{cs} c_1=-\frac12\,,\qquad\text{and}\qquad c_2=-\frac38\,.\ee 
Background diffeomorphism invariance tells us that the linearised curvatures accompanying $\bar{\beta}$ and $\bar{\gamma}$ simply become full curvatures (by \eqref{deflinearisedcurvature} they absorb one power of $\kappa$) but, as discussed in sec. \ref{sec:onelooptwoptagone}, the appearance of the linearised quantum curvatures in \eqref{K1ag1-2pt} is accidental, so it is more appropriate for the $\beta$ and $\gamma$ pieces to appear with their separate parts covariantised, following \eqref{curvatures}. Even though all these parameters are known, and that includes $\tau$ \cite{Gibbons:1978ac,Goroff:1985th}, we leave them general when we match to the three-point one-loop results, as extra checks on the formalism. 

The remaining eleven $\alpha_i$, twenty-seven $b_i$, and twenty-one $d_i$, are genuinely free parameters to be determined. The schematic representation for the $d_i$ terms means that one sums over  the vertices with coefficients $d_i$, these vertices being the twenty-one linearly independent combinations of two spacetime derivatives and one $c^*_\alpha$, $c^\beta$, and $h_{\gamma\delta}$. We ensure independence under integration by parts by taking as representatives  those vertices where $c^*_\alpha$ is undifferentiated. Since the $d_i$ terms are already three-point vertices, as are the $b_i$ terms, background covariantisation is ignored there. For the same reason, we actually do not need diffeomorphism invariant expressions for the $  \alpha_8,\cdots,  \alpha_{11}$ terms, whilst in the other $\alpha_i$ terms we actually only need the linearised background curvature.

The sum over $b_i$ vertices is defined in the same way as for the $d_i$ vertices, except that all terms involving $\partial_\alpha h^{*\alpha\beta}$ are discarded, and likewise any two vertices should be considered equal if they only differ by such terms on using integration by parts. (This can be implemented  straightforwardly by deriving the vertices in momentum space.) The reason for this restriction is because at the three-point level, vertices containing       $\partial_\alpha h^{*\alpha\beta}$ are already accounted for in the $d_i$ sum. As in the discussion in sec. \ref{sec:onelooptwoptagone}, this is a consequence of the fact that we can add an $s_0$-exact part to $K_{1/1}$ without altering $S_{1/1}$, \cf \eqref{oneloopgensoln}.
At the three-point level we can add $(Q+Q^-) (c^* h^2\partial)$, but $Q^-$ generates the $\partial_\alpha h^{*\alpha\beta}$ terms while $Q$ maps onto combinations in the $d_i$ sum that contain $\partial_{(\alpha}c_{\beta)}$.
Finally, for the same reason we do not want a free parameter for the combination 
\be -\frac{\kappa}{\sqrt{\bar{g}}} s_0 \left( \ph^* h^{*\mu\nu}h_{\mu\nu} \right) = \bar{R}h^{*\mu\nu}h_{\mu\nu}+2\ph^*\bar{R}^{\alpha\beta}h_{\alpha\beta}-2\bar{R}\ph^*\ph -2\frac{\kappa}{\sqrt{\bar{g}}}\ph^*h^{*\mu\nu}\bar{\nabla}_\mu c_\nu\,.\ee
The last three terms on the right hand side appear in our parametrisation, but this is why the first term is missing from it.

Although the resulting parametrisation is long, it is a dramatic reduction compared to the thousands of vertices from the Feynman diagram calculation, and therefore in fact the parameters are vastly overdetermined. That we nevertheless find a consistent solution for all vertices is thus a highly non-trivial verification of the formalism.

Matching to just the (antighost level zero) pure background $\bar{h}^3$ vertices, we reproduce well-known results: we confirm that the pure background curvature-squared terms at the two-point level, \cf \eqref{oneloopzerotwopt}, are covariantised to full background curvatures, as is in fact clear here from our $K^1_{1/1}$ \eqref{K11}, and confirm that the remaining part is the Gauss-Bonnet term given in \eqref{GB}. In this way we reaffirm the $\bar{\beta}$ and $\bar{\gamma}$ values from \eqref{betagammavals} and also find 
\be \tau=\frac{53}{90} \,,\ee
in agreement with previous calculations \cite{Gibbons:1978ac,Goroff:1985th}. 

One can determine all the coefficients in $K^1_{1/1}$ by matching to antighost level zero vertices, up to several vertices parametrised by $c_1$.  In fact just using the $h^2\bar{h}$ and $h^3$ vertices is sufficient to determine all that can be found at this level, but we matched also to $\bar{h}^2h$ vertices to verify the result and further confirm consistency. The $K^2_{1/1}$ parameters cannot of course be determined by matching to antighost level zero vertices, because the lowest antighost level it generates is level one, via $Q^-K^2_{1/1}$, while $c_1$ and some vertices in the $b_i$ sum also remain undetermined  because in $K^1_{1/1}$ at the three-point level they can be collected  into $\half c_1 Q^-(c^{*\nu}\bar{\nabla}_\nu\ph)$. 

Now all the parameters in $K^2_{1/1}$, and $c_1$, can be (over)determined by matching to the full set of level-one three-point Feynman diagrams with topology of fig. \ref{fig:threepoint}, \ie such that one external leg is a ghost $c^\mu$, one external leg is $h^{*\alpha\beta}$ and the remaining leg is $h$ or $\bar{h}$. In this way we recover the previously stated values for $c_1$, $c_2$, $\bar{\beta}$, $\bar{\gamma}$, $\beta$ and $\gamma$, and determine that
\begin{align}
        \alpha_1=& -\frac{1}{8}\,,\quad \alpha_2=-\frac{1}{24}\,,\quad \alpha_3= \frac{161}{120}\,,\quad \alpha_4= \frac{1}{120}\,,\quad   \alpha_5=-\frac{3}{4}\,, \quad   \alpha_6=-\frac{7}{15}\,, \nonumber \\
         \quad   \alpha_7=& \frac{19}{60}\,,\quad   \alpha_8=-\frac{1}{6}\,,\quad   \alpha_9=-\frac{1}{12}\,, \quad
          \alpha_{10}=-\frac{4}{15} \,,\quad   \alpha_{11}=-\frac{1}{6} \,,
\end{align}
and also the $b_i$ and $d_i$ parameters as given below:
\beal
      \sum_{i=1}^{27} & b_i  \left(h^{*}h^2\partial^2\right)_i  =   \frac{5}{12}  h^{*\mu\nu}\varphi \partial^2_{\mu\nu}\varphi    
         -\frac{13}{160}   h^{*\mu\nu}\partial^2_{\mu\nu}  h^\beta{}_\alpha  h^\alpha{}_\beta    + \frac{1}{4}   h^{*\mu\alpha}\left(\partial^\nu  h_{\alpha\nu}\partial_\mu\varphi      
         -\partial_\mu\partial^\nu  h_{\alpha\nu}\varphi \right) \nn  \\
       &  +\frac{61}{240}   h^{*\mu\alpha} \left( \partial_\mu  h_{\alpha\nu}\partial^\nu \varphi   - h_{\alpha\nu}\partial^\nu\partial_\mu\varphi  \right)         
   + \frac{7}{80}  h^{*\mu\alpha} \left( \partial_\mu  h_{\beta\nu} \partial^\nu  h^\beta{}_\alpha  - h_{\beta\nu}  \partial^\nu \partial_\mu  h^\beta{}_\alpha \right)\nn\\
 &  -\frac{61}{240}  h^{*\mu\alpha}\left(\partial_\nu  h_{\beta}{}^{\nu}\partial_\mu  h^\beta{}_\alpha -\partial^2_{\mu\nu} h_{\beta}{}^\nu  h^\beta{}_\alpha \right)  
+\frac{13}{60}\ph^*\partial^\alpha h_{\beta\nu}\partial^\nu h^\beta{}_\alpha
+\frac{43}{60} \varphi^*   h_{\beta\nu} \partial^\nu \partial^\alpha  h^\beta{}_\alpha  \nn \\
  &       +\frac{77}{120}  \varphi^* \partial^\nu  h_{\beta\nu}\partial^\alpha   h^\beta{}_\alpha     -\frac{53}{60}  \varphi^*   h_{\alpha\nu} \partial^\nu \partial^\alpha \varphi     -\frac{17}{10}  \varphi^* \partial^\nu  h_{\alpha\nu}\partial^\alpha  \varphi 
         -\frac{3}{10}  \varphi^*  \varphi \square \varphi     -\frac{11}{60}  \varphi^*\varphi \partial^2_{\alpha\nu} h^{\alpha\nu}    \nn\\  
         & +\frac{9}{40}  \varphi^*   h^\alpha{}_\beta \square  h^\beta{}_\alpha 
         +\frac{14}{15}  \varphi^* \partial_\nu  \varphi  \partial^\nu \varphi    
         -\frac{11}{80} \varphi^* \partial_\nu  h^\alpha{}_\beta \partial^\nu  h^\beta{}_\alpha   -\frac{131}{240}   h^{*\mu\nu} \partial^\alpha  h_{\mu\nu}  \partial_\alpha \varphi  \nn\\
     &           -\frac{1}{4}  h^{*\mu\nu}  h^\alpha{}_\mu\partial^2_{\alpha\beta}  h^\beta{}_\nu    
         -\frac{1}{12}   h^{*\mu\nu} \partial_\alpha  h^\alpha{}_\mu \partial_\beta  h^\beta{}_\nu    
         -\frac{27}{80}   h^{*\mu\nu} \partial_\beta  h^\alpha{}_\mu \partial_\alpha  h^\beta{}_\nu     +\frac{17}{80}   h^{*\mu\nu} \partial_\alpha  h_{\mu\beta} \partial^\alpha  h^{\beta}{}_\nu  \nn\\
&         +\frac{7}{80}  h^{*\mu\nu} \partial^2_{\alpha\beta}  h_{\mu\nu}  h^{\alpha\beta}     
         -\frac{1}{2}   h^{*\mu\nu} \square  h_{\mu\beta}   h^{\beta}{}_\nu   
         +\frac{37}{80}   h^{*\mu\nu} \partial^\alpha  h_{\mu\nu} \partial^\beta  h_{\alpha\beta}       
        -\frac{1}{3}  h^{*\mu\nu}   h_{\mu\nu}\square  \varphi  \nn\\
   &      +\frac{1}{3}  h^{*\mu\nu}   h_{\mu\nu} \partial^2_{\alpha\beta}  h^{\alpha\beta}     +\frac{11}{24}   h^{*\mu\nu}\square   h_{\mu\nu}  \varphi \,.    
\eeal
\beal
   \sum^{21}_{i=1} & d_i \left(c^*ch\partial^2\right)_i  =
         \frac{1}{12}  c^{*\mu} \partial^2_{\mu\nu} c^{\nu} \varphi       
         -\frac{121}{480} c^*_\mu \partial^\mu c^{\nu}\partial_\nu \varphi     
         +\frac{61}{480}  c^*_\mu \partial^\mu c^{\nu} \partial_\alpha  h^\alpha{}_\nu     
         -\frac{11}{24}  c^{*\mu} \partial^2_{\mu\alpha} c^{\nu}  h^\alpha{}_\nu     \nn\\ &  
         -\frac{1}{3} c^{*\mu} c^{\nu} \partial^2_{\mu\nu} \varphi     
          +\frac{1}{6} c^{*\mu} c^{\nu} \partial^2_{\alpha\mu}  h^\alpha{}_\nu     
         -\frac{1}{24}  c^*_\mu \partial_\nu c^{\nu} \partial^\mu \varphi        
         -\frac{101}{480} c^*_\mu \partial_\alpha c^{\nu}\partial^\mu  h^\alpha{}_\nu        
         -\frac{1}{8} c^*_\alpha\partial^2_{\mu\nu} c^{\nu}  h^{\alpha\mu}      \nn\\ & 
         -\frac{119}{480} c^*_\alpha \partial^\mu c^{\nu}\partial_\nu  h^\alpha{}_\mu      
          +\frac{1}{12}  c^*_\alpha c^{\nu} \partial^2_{\mu\nu}  h^{\alpha\mu}   
          +\frac{1}{8}  c^*_\alpha \partial_\nu c^{\nu}\partial^\mu  h^\alpha{}_\mu     
         -\frac{301}{480}  c^*_\alpha\partial^\nu c^{\alpha}\partial_\nu \varphi      
          +\frac{1}{4}  c^*_\alpha c^{\alpha} \square \varphi      \nn\\ &
          +\frac13c^*_\alpha\Box c^\alpha\ph  
          -\frac{1}{12} c^*_\alpha c^{\alpha}\partial^2_{\mu\nu} h^{\mu\nu}     
          +\frac{27}{160}  c^*_\alpha\square c^{\mu}  h^\alpha{}_\mu      
         -\frac{239}{480}  c^*_\alpha \partial_\nu c^{\mu} \partial^\nu  h^\alpha{}_\mu       
         -\frac{1}{4}  c^*_\alpha c^{\mu} \square  h^\alpha{}_\mu     \nn\\ &
         +\frac{7}{160}  c^*_\alpha \partial^2_{\mu\nu} c^{\alpha}  h^{\mu\nu}       
          +\frac{241}{480}  c^*_\alpha \partial^\nu c^{\alpha}\partial^\mu  h_{\mu\nu}     \,.
\eeal

Since the above provides us with the full expression for $K_{1/1}$ up to the three-point level, we get as a bonus the full expression up to three-point level for the antighost level-two counterterm, without having to compute it from Feynman diagrams, since it is given by $S^2_{1/1} = QK^2_{1/1}$. This completes the explicit calculation of all off-shell one-loop divergences with up to three (anti)fields.

\subsection{Two-loop double-pole two-point graviton counterterms}
\label{sec:twoloops}

Now as advertised we use the one-loop counterterm diagrams, illustrated in fig. \ref{fig:twopointcountertermdiags}, to compute the two-loop $1/\eps^2$ counterterm via the RG relation \eqref{countertermkey}. We limit ourselves to the two-point diagrams at antighost level zero, \ie with either a quantum or background graviton external leg. This is already enough for a non-trivial explicit test of the second order canonical expansion relation \eqref{Stwogensol}.

\ignore{
\begin{figure}[ht]
\centering
$$
\begin{gathered}
\begin{tikzpicture}
  \begin{feynman}
    \vertex (a) ;
    \vertex [right=of a] (b) ;
     \vertex [right=of b] (c);
      \vertex [right=of c] (d);
     \vertex at ($(b)!0.5!(c)!0.5cm!90:(c)$) [crossed dot] (t) {};
\diagram* {
      (a) -- [plain] (b), (b) -- [plain, half right] (c),
      (c) -- [plain] (d), (b)
-- [plain, bend left] (t), (t)
-- [plain, bend left] (c),
    }; 
  \end{feynman}
\end{tikzpicture}
\end{gathered}
\qquad
\begin{gathered}
\begin{tikzpicture}
  \begin{feynman}
    \vertex (a) ;
    \vertex [right=of a,crossed dot] (b) {};
     \vertex [right=of b] (c);
      \vertex [right=of c] (d);
\diagram* {
      (a) -- [plain] (b), (c) -- [plain, half left,looseness=1.2] (b),
      (c) -- [plain] (d), (c)
-- [plain, half right,looseness=1.2] (b),
    }; 
  \end{feynman}
\end{tikzpicture}
\end{gathered}
$$
\caption{RG relates the $1/\eps^2$ pole in the two-loop two-point counterterm vertices to one-loop counterterm vertices, represented by the crossed circles, via one-loop counterterm diagrams with the above topologies.}
\label{fig:twopointcountertermdiags}
\end{figure}
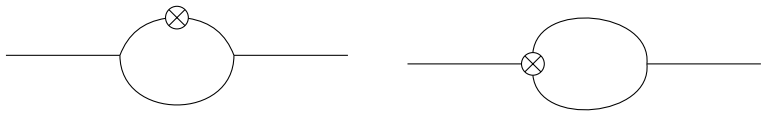}

For the first diagram in fig. \ref{fig:twopointcountertermdiags}, we need the one-loop two-point counterterm vertices with purely quantum legs. They are given by  \eqref{onelooponetwopt} and the first two terms in \eqref{oneloopzerotwopt} for ghosts and graviton  respectively. (In the former case we need to shift to gauge fixed basis using relation \eqref{gf to gi 1} at the linearised level.) For the second diagram we need the one-loop three-point counterterm vertices with two quantum legs  and either an external $h_{\alpha\beta}$ or $\bar{h}_{\alpha\beta}$. These can be ported directly from intermediate results created as a side-product of the computation reported in the previous subsection. Alternatively, they can be generated by evaluating $s_0K_{1/1}$ using the explicit expressions given there. (As expected the topological counterterm \eqref{GB} can be disregarded since it makes no contribution to the Feynman integrals.)

The result we find is that for two-point vertices:
\begin{align}
    S^0_{2/2}= -\frac12\frac{\kappa^4\mu^{4\eps}}{(4\pi)^4\eps^2}\int_x \Bigg\{ &\frac{11}{36} \Bar{R}^{(1)}\square \Bar{R}^{(1)} +\frac{5}{72}\Bar{R}^{(1)\mu\nu}\square \Bar{R}^{(1)}_{\mu\nu} - \frac{469}{3600} R^{(1)}\square R^{(1)}+\frac{79}{200} R^{(1)\mu\nu}\Box R^{(1)}_{\mu\nu} \nonumber \\
    +& \frac{781}{3600} \Bar{R}^{(1)}\square R^{(1)}+\frac{53}{150}\Bar{R}^{(1)\mu\nu}\square R^{(1)}_{\mu\nu}-\frac{31}{720}\left( \Bar{R}^{(1)} + R^{(1)}\right) \square^2 \varphi \Bigg\}\,,\label{Stwotwo}
\end{align}
where the overall factor of $-\half$ is the conversion \eqref{countertermkey} from the double-pole in fig. \ref{fig:twopointcountertermdiags} to the two-loop counterterm $S_{2/2}$. As we will see this result passes a highly non-trivial consistency check in that it satisfies the second order canonical transformation relation \eqref{Stwogensol}. 
As far as we know the above result has not appeared in the literature before, except for the one term: $\Bar{R}^{\mu\nu}\bar{\square} \Bar{R}_{\mu\nu}$ \cite{Goroff:1985sz}. However this was quoted there as part of some partial results that unfortunately contain an error \cite{Solodukhin:2020vuw}. Nevertheless comparing the coefficients for this one term, we find that they agree up to a factor of half,  see app. \ref{app:comparisons}.

Recall that the one-loop level-zero two-point result \eqref{oneloopzerotwopt} can be written entirely in terms of linearised curvatures \eqref{curvatures} and is thus invariant under (linearised) diffeomorphisms, in particular also for the fluctuation field $h_{\mu\nu}$. This latter invariance is a consequence of invariance under the linearised BRST charge $Qh_{\mu\nu}=\partial_{(\mu}c_{\nu)}$. Recall also from sec. \ref{sec:onelooptwoptagone} that this property is actually an accident of Feynman DeDonder gauge. The presence of the $\Box^2\ph$ term above shows that at two loops, one's luck runs out and this property is violated. As is evident from the form of this last term, it just corresponds to inserting another $\Box$ into the unique one-loop $Q$-invariance-breaking possibility \eqref{oneloopQbreaking}.

In the remainder of this subsection we will show that the double-pole \eqref{Stwotwo} corresponds to a canonical transformation taken to second order, \ie can be expressed as in \eqref{Stwogensol}: 
\be \label{Stwogensoll} S_2 = \frac12(S_1,K_1)+s_0K_2\,.\ee
Actually recall that this expression follows from the non-linear CME relation $s_0S_2=-\frac12(S_1,S_1)$, \viz \eqref{StwoCME}, on assuming that $S_1$ is given only by the exact piece $s_0K_1$, whereas the one-loop solution \eqref{oneloopgensoln} also contains the Gauss-Bonnet term \eqref{GB}. However since the latter is topological it makes no contribution to the antibracket and thus \eqref{Stwogensoll} is indeed the correct solution. 

From sec. \ref{sec:threepoints}, it is clear that $(S_1,S_1)$ cannot vanish at the three-point level, and thus the non-linear CME relation itself is highly non-trivial. However for the two-point vertices $(S_1,S_1)$ in fact does vanish. This is straightforward to see by inspection since for the two-point vertices we only have the pure curvature antighost level zero part, $S^0_{1/1}$, as given in \eqref{oneloopzerotwopt}, and the antighost level one part, $S^1_{1/1}$, as given in \eqref{onelooponetwopt}. But substituting these into $(S_{1/1},S_{1/1})$ the net effect is to replace $h_{\mu\nu}$ in a `quantum curvature' by either $\partial^3_{\mu\nu\alpha}c^\alpha$ or $\Box\partial_\mu c_\nu$ (up to some coefficient of proportionality), causing the result to vanish since both of these are pure gauge. 

Thus the non-linear CME relation \eqref{StwoCME} only implies that the two-point vertex in $S_2$ is $s_0$-closed. The problem is that the two-point level is to a certain extent degenerate. A related point is that if we take the action only to have an antighost level zero piece, and take this to be any product of linearised curvatures, that is any one of the terms in $S^0_{1/1}$ of \eqref{oneloopzerotwopt}, then this action is $s_0$-closed at the two-point level since the linearised quantum curvatures are invariant under linearised diffeomorphisms. Nevertheless as we will see, this test is still non-trivial because 
although at the level of two-point vertices $\frac12(S_1,K_1)$ in the general solution \eqref{Stwogensoll} is $s_0$-closed, it is not $s_0$-exact.

Specialising \eqref{Stwogensoll} to antighost level zero and divergences we have
\be S^0_{2/2} = \frac12(S^0_{1/1},K^1_{1/1})+Q^-K^1_{2/2}\,.\label{Stwotwogensol}\ee
Substituting $\int_x h^{*\mu\nu}\partial^2_{\mu\nu}\ph$ for $K^1_{1/1}$ into the antibracket, we see that it vanishes for the same reasons as above. Therefore the \eqref{K1ag12-2pt} part of $K_{1/1}$ makes no contribution. Since the remaining part of $K_{1/1}$, \viz \eqref{K1ag1-2pt}, is made of linearised curvatures, we see that the antibracket contributes terms with linearised curvatures only. Explicitly, we find
\besp \label{twoloopabracket}
    \frac{1}{2}(S^0_{1/1},K^1_{1/1})=\frac{\kappa^4\mu^{4\eps}}{(4\pi)^4\eps^2}\int_x\Big\{ -\frac{1}{2}\Bar{\beta}(\beta +\Bar{\beta})\bR_{\mu\nu}\square \Bar{R}^{(1)\mu\nu}-\beta^2 \R_{\mu\nu} \square R^{(1)\mu\nu}\\
 +\frac{1}{2}(3\gamma^2+2\beta \gamma+\beta^2) \R\square \R 
+\frac{1}{4}(3\Bar{\gamma}^2+3\Bar{\gamma}\gamma+2\Bar{\beta}\Bar{\gamma}+\Bar{\beta}\gamma+\Bar{\beta}^2+\beta\Bar{\gamma}+\beta\Bar{\beta})\bR\square\bR\\
-\frac{1}{2}\beta(\beta + 3 \Bar{\beta})\R_{\mu\nu}\square \Bar{R}^{(1)\mu\nu}
    +\frac{1}{4}(9\gamma\Bar{\gamma}+3\gamma^2+3\Bar{\beta}\gamma +3\beta \Bar{\gamma}+2\beta\gamma +3\beta \Bar{\beta}+\beta^2)\R\square \bR\Big\}\,,
\eesp
where recall that the parameters were determined as in \eqref{betagammavals}. Now this cannot come from an $s_0$-exact expression because if it did, we could write it as $Q^-K^1$ for some $K^1$. We can check if this is so by using the same rule discussed in sec. \ref{sec:onelooptwoptagzero}, \ie from \eqref{Q-h*linearised} we know that this would imply that the coefficient of the mixed terms above must be equal to the sum of the coefficients of the equivalent pure quantum and pure background pieces. It is easy to see that this does not work. Similarly one can verify that the curvature terms in \eqref{Stwotwo} do not sum to something that is $Q^-$-exact. 

But according to \eqref{Stwotwogensol}, on subtracting \eqref{twoloopabracket} from \eqref{Stwotwo} we should be left with a  $Q^-$-exact piece. We have already seen that this is true of the non-covariant term, the last term, in \eqref{Stwotwo}. The remaining parts are pure curvature terms and must thus have the parametrisation \eqref{K1ag1-2pt} except with an extra $\Box$ inserted (and different coefficients), up to some $Q^-$-exact remainder, $Q^-R\in K^1_{2/2}$ (which does not contribute to \eqref{Stwotwogensol} because $Q^-$ is nilpotent). Matching to the above results, we find that this is indeed the case and thus we derive $K^1_{2/2}$ at the two-point level in the form
\besp K^1_{2/2} \ = \ \frac{\kappa^4\mu^{4\eps}}{(4\pi)^2\eps^2}\int_x \Big\{ \frac{877}{28800} h^{*\mu\nu}\Box R^{(1)}_{\mu\nu}
+\frac{71}{1800}\varphi^*\Box R^{(1)}\\
+\frac{361}{28800} h^{*\mu\nu}\Box \Bar{R}^{(1)}_{\mu\nu}+\frac{2719}{14400}\varphi^*\Box \Bar{R}^{(1)} -\frac{31}{1440}\ph^*\Box^2\ph\Big\} + Q^-R\,.\label{K2}\eesp
Like in \eqref{K1ag1-2pt}, the remainder term $Q^-R$ has $\partial_\alpha h^{*\alpha\beta}$ as a factor. It could also be derived by matching to the two-loop double-pole level-one counterterm diagrams, and they can be computed using the results we have already obtained. However the above form for $K_{2/2}$ is sufficient for our purposes.

\section{Generalised beta functions and why they are not finite}
\label{sec:solodukhin}

In this final section we comment on some ideas for generalised $\beta$-functions, where the field is taken to play the r\^ole of a collection of couplings. The key idea is to exploit relations that follow from assuming that these $\beta$-functions are finite. Unfortunately this assumption is incorrect. We explain why natural generalisations that respect the BRST symmetry also fail to work.

Inspired by ref. \cite{Kazakov:1987jp} and its many follow-ups \eg \cite{Kazakov:2019wce,Kazakov:2022pkc}, which themselves are inspired by refs. \cite{Friedan:1980jf,Friedan:1980jm,Alvarez-Gaume:1981exa}, the main proposal of ref. \cite{Solodukhin:2020vuw} consists of two key steps. The first key step is to allow for a non-linear renormalisation of the metric, replacing $g_{\mu\nu}$ in the Einstein-Hilbert term of the classical action \eqref{minimal classical action} with a bare metric $g^0_{\mu\nu}$ which is then expanded as 
\be\label{grenorm} g^0_{\mu\nu}(x) = g_{\mu\nu}(x) + \sum_{k=1} \frac1{\varepsilon^k}\, \mathfrak{g}^k_{\mu\nu}(x)\,.  \ee
The $\mathfrak{g}^k_{\mu\nu}$ are assumed to be local diffeomorphism covariant combinations constructed from covariant derivatives and curvatures using the renormalised metric $g_{\mu\nu}$. With this assumption, the proposal only applies to non-linear renormalisation of the background metric.

In ref. \cite{Solodukhin:2020vuw} the $\mu$ dependence in \eqref{grenorm} is simplified to an overall multiplicative $\mu^{-2\eps}$ on the right hand side,  by taking the mass dimensions to be  $[g^0_{\mu\nu}]=-2\eps$, while $[g_{\mu\nu}]=0$ and $[\kappa]=-1$ (also in $d$ dimensions). 
However the same physics can be arrived at by including $\mu$ in the more conventional way, as we do in this paper. Thus our metrics are taken to be  dimensionless, while $[\kappa]=-1+\eps$. Then by dimensions, the $\mathfrak{g}^k_{\mu\nu}$ are forced to have explicit dependence on $\mu$, \cf sec. \ref{sec:RG} and sec. \ref{sec:Loops}. In fact the $\ell$-loop contribution is constructed from $2\ell$ covariant derivatives, rendered dimensionless by the factor $(\kappa\mu^{-\eps})^{2\ell}$.

A renormalisation of form \eqref{grenorm} can provide all the covariant counterterms in the bare action that vanish on the equations of motion. For example the purely background metric counterterms (in Feynman -- De Donder gauge) are \cite{tHooft:1974toh}, \cf \eqref{oneloopzerotwopt} and below \eqref{cs},
\be 
\label{tHooftVeltman}
S_1 = \frac{\mu^{-2\eps}}{(4\pi)^2\eps}\,\int_x \sqrt{\bar{g}}\left( \frac1{120}\bar{R}^2+\frac7{20}\bar{R}_{\mu\nu}^2\right)\,.
\ee
These counterterms can be generated by defining
\be 
\label{grenormeg}
\bar{g}^0_{\mu\nu} = \bar{g}_{\mu\nu} +\frac{\kappa^2\mu^{-2\eps}}{(4\pi)^2\eps} \, \bar{\mathfrak{g}}^1_{\mu\nu}\,,\qquad\text{where}\qquad 
\bar{\mathfrak{g}}^1_{\mu\nu} = \frac7{40}\bar{R}_{\mu\nu}+\frac{11}{120}\bar{g}_{\mu\nu}\bar{R}
\ee
(where, from here on, we make explicit the ${\kappa\mu^{-\eps}}/{(4\pi)}$ dependence in $\bar{\mathfrak{g}}^k_{\mu\nu}$).

Now by insisting that the bare metric is independent of $\mu$, and differentiating both sides with respect to $\mu$, one obtains a kind of generalised ``beta function'', $\beta_{\alpha\beta} 
= \mu\partial_\mu\, g_{\alpha\beta}$ for the renormalised metric (non-linear wavefunction renormalisation might be a better term). For the above example, from \eqref{grenormeg}, we have for the background metric to one loop,
\be 
\label{betabackone}
\bar{\beta}_{\mu\nu} = 2\,\frac{\kappa^2\mu^{-2\eps}}{(4\pi)^2}\, \bar{\mathfrak{g}}^1_{\mu\nu}\,.
\ee
The second key step is actually implicit in ref. \cite{Solodukhin:2020vuw}. It is the assumption that such generalised beta functions are finite in the limit $\eps\to0$. We have just seen that this is trivially true at one loop, but at higher loops this is a powerful assumption. Just as with the usual beta functions in a renormalisable theory, the one-loop result would then be enough to determine the leading pole $1/\eps^\ell$ at each loop order $\ell$ {without computing any more Feynman diagrams}. To see this in our example, assume we already know the leading two-loop purely background counterterm and have chosen $\bar{\mathfrak{g}}^2_{\mu\nu}$ to generate it via
\be 
\label{grenormex2}
\bar{g}^0_{\mu\nu} = \bar{g}_{\mu\nu} +\frac{\kappa^2\mu^{-2\eps}}{(4\pi)^2\eps}\, \bar{\mathfrak{g}}^1_{\mu\nu}+\frac{\kappa^4\mu^{-4\eps}}{(4\pi)^4\eps^2}\, \bar{\mathfrak{g}}^2_{\mu\nu}\,,
\ee
where the prefactor follows because $\bar{\mathfrak{g}}^2_{\mu\nu}$ will be formed from four background covariant derivatives. Then cancellation of the $1/\eps$ single-pole in $\bar{\beta}_{\mu\nu}$ tells us that 
\be \bar{\mathfrak{g}}^2_{\alpha\beta} = \frac{4\pi^2\mu^{2\eps}}{\kappa^2}\, \mu\partial_\mu\bar{\mathfrak{g}}^1_{\alpha\beta}[\bar{g}] \,,\ee
Applying the Leibniz  rule and using \eqref{betabackone}, we see that $\bar{\mathfrak{g}}^2_{\alpha\beta}$ should in fact be computable simply by applying a first order shift of the background metric on the one-loop result:
\be \bar{\mathfrak{g}}^2_{\alpha\beta} = 
\delta\bar{\mathfrak{g}}^1_{\alpha\beta}[\bar{g}]\,,\qquad\text{where}\qquad  \delta \bar{g}_{\mu\nu} = \frac12 \,\bar{\mathfrak{g}}^1_{\mu\nu}\,. \ee

Unfortunately this does not work as can be verified explicitly at the two-point level by using the pure background terms from \eqref{Stwotwo} (for higher order see the discussion below that equation). The reason is that the second key step, the assumption that these generalised beta functions are finite, is incorrect. In the original incarnation as applied to the target metric of the two-dimensional sigma model \cite{Friedan:1980jf,Friedan:1980jm,Alvarez-Gaume:1981exa}, it was correct, because the target metric actually represents an infinite set of couplings. But applied to the fields themselves, as in the proposal of ref. \cite{Solodukhin:2020vuw}, it is not correct. 

The obstruction to finiteness of $\bar{\beta}_{\mu\nu}$ shows up most clearly in the gauge fixing. The result \eqref{tHooftVeltman} is derived using De Donder gauge \eqref{F}. Clearly the transformation \eqref{grenormex2} alters the gauge \eqref{F} (by a divergent amount). That is a problem because the Legendre effective action is not the same in different gauges except on shell. But $\bar{\mathfrak{g}}^2_{\mu\nu}$ in \eqref{grenormex2} has been chosen to cancel a part that only exists off shell. 

In fact let us now recall that counterterms are required that depend on all combinations of the fields, in particular the  quantum fields, as we have seen. In the background field method it is possible to work exclusively with diagrams that have only external background field legs (as in \eg \cite{Goroff:1985th}). However even if we do not explicitly track the value of counterterms that cancel divergences in vertices involving quantum fields, they must be there in practice because they cancel sub-divergences in higher loops, and higher loop divergences are local as required only if all these sub-divergences have been cancelled \cite{tHooft:1973wag,Caswell:1981ek,Chase:1982sf}, as we recalled in sec. \ref{sec:RG}. 

Then as we saw in sec. \ref{sec:General form of divergences in the Legendre effective action}, the `new' divergences at each loop order are $s_0$-closed. Those that vanish on the equations of motion, are $s_0$-exact and correspond to infinitessimal canonical transformations \eqref{canonsmall} between the antifields and quantum fields. As we proved there, and also verified in sec. \ref{sec:Loops}, there is no separate purely background renormalisation. What happens instead is that purely background counterterms also get absorbed by these canonical transformations. 
This extends to the non-linear terms that appear beyond one loop order. For example we saw that the leading (\ie double-pole) counterterm at two loops, \eqref{Stwotwogensol}, also involves carrying the one-loop canonical transformation to second order, as we saw in sec. \ref{sec:canonsecond}.

Now it is clear that if the proposal of \cite{Solodukhin:2020vuw} is going to work, it should apply not to the background metric, but to the antifields and quantum fields. Indeed the second-order canonical transformation $\delta\phi^{(*)}$ given in eqn. \eqref{canonsecond}, is  the correct non-linear transformation between bare (anti)fields 
\be \phi_0^{(*)} = \phi^{(*)}+\delta\phi^{(*)}\ee 
and renormalised (anti)fields $\phi^{(*)}$, such that it will generate through Taylor expansion \eqref{STaylor} of the classical action, all the required counterterms that vanish on shell, up to two loops.\footnote{The Jacobian for this local transformation vanishes in dimensional regularisation, recall below \eqref{Wardq}.} 

Independence of $\phi_0^{(*)}$ on $\mu$, then implies the generalised beta functions 
\be \beta^A[\phi,\phi^*]= \mu\partial_\mu \phi^A\qquad\text{and}\qquad \beta^*_A[\phi,\phi^*] = \mu\partial_\mu \phi^*_A\,. \ee
Following the previous argument, if we assume that these beta functions are finite, we can derive $K_2$ from $K_1$ without computing Feynman diagrams. Once again we can check this idea explicitly using the results for $K_1$ from sec. \ref{sec:onelooptwopt}. It turns out that it implies that at the two-point level $K_2$ must vanish. But from \eqref{K2} this is incorrect. In fact, irrespective of the details, this proposal cannot work because the $K_1$ terms just furnish linearised curvatures for $K_2$, whereas $K_2$ has the explicitly non-covariant piece -- the last term under the integral in \eqref{K2}. Again, the mistake in this reasoning is the assumption that the generalised beta functions are finite. 

To see why they cannot be finite, note that the partition function \eqref{partnfn} now takes the form
\begin{equation}
    \mathcal{Z}[J,\phi^*] = \int\!\! \mathcal{D} \phi\, \mathrm{e}^{-S[\phi_0,\phi^*_0] + \phi^A J_A} \,,
\end{equation}
Here the bare antifields are responsible for generating all the counterterms that vanish on shell, via canonical transformations  \eqref{canonsecond}, whilst $S$ itself contains the counterterms for cohomologically non-trivial pieces which depend only on the total metric, such as the topological term \eqref{GB} at one loop, and the Goroff-Sagnotti term \cite{Goroff:1985sz}
\be \label{GS}
S_2\ \ni \ \frac{209}{5760}\frac{\kappa^2\mu^{-2\eps}}{(4\pi)^4\eps}\int_x\!\sqrt{g}\,R_{\alpha\beta}^{\phantom{\alpha\beta}\gamma\delta}R_{\gamma\delta}^{\phantom{\alpha\beta}\epsilon\zeta}R_{\epsilon\zeta}^{\phantom{\alpha\beta}\alpha\beta}
\ee
at two loops. All Green's functions are then finite (in particular this is so for the Legendre effective action, which is a functional of the classical fields $\Phi^A$ and the renormalised antifields $\Phi^*_A=\phi^*_A$). However for the operators that vanish on shell, we are now attributing $\mu$ dependence to the renormalised (anti)fields $\phi^{(*)}$ rather than renormalised couplings $c_\ell^i$ as before. Unfortunately $\mu$-independence of the bare action $S[\phi_0,\phi^*_0]$ then implies that $\beta^A$ cannot be finite since:
\be \mu\partial_\mu \mathcal{Z}[J,\phi^*] = \int\!\! \mathcal{D} \phi\ \beta^A[\phi,\phi^*]J_A\ \mathrm{e}^{-S[\phi_0,\phi^*_0] + \phi^A J_A} \label{betadiverge}\,.\ee
Indeed the left hand side is finite by construction, but the right hand side involves the insertion of $\beta^A$ which is local and non-linear in renormalised quantum fields. The insertion of such terms generates new divergences, and the only way they can be cancelled is if in fact $\beta^A$ already contains precisely the right divergences to cancel them.

\section{Discussion and Conclusions}
\label{sec:conclusions}

Off-shell counterterms in quantum gravity, defined perturbatively as an effective theory about a background metric $\bar{g}_{\mu\nu}$, are invariant under background diffeomorphisms, BRST, and the RG. In this paper we have drawn out some of the consequences of the way these symmetries are interwoven with each other. 

In particular we have shown in sec. \ref{sec:General form of divergences in the Legendre effective action} that at each new loop order the new divergences, those that are annihilated by the total classical BRST charge $s_0$, can be characterised as being either diffeomorphism invariant functionals of the total metric $g_{\mu\nu}$ which do not vanish on the classical equations of motion (\ie do not vanish when $G_{\mu\nu}=0$) or as $s_0$-exact functionals which are thus first order canonical transformations of the antifields and quantum fields (\cf sec. \ref{sec:canonsecond}). In particular we show that there are no separate purely background field divergences. Then it follows that those background field terms that do not vanish on the equations of motion $\bar{G}_{\mu\nu}=0$, are part of the diffeomorphism invariant functionals of the total metric, whilst those that do vanish on the equations of motion are renormalised by reparametrising the quantum fluctuation $h_{\mu\nu}$ as part of the canonical transformations. The background metric itself is never renormalised.

By adding the antifield sources for BRST transformations, we keep track of the deformations of the BRST algebra induced by renormalisation. These appear as part of the $s_0$-exact counterterms. Whilst the Zinn-Justin/CME equation is preserved at each loop order $\ell$ for both the bare action and the Legendre effective action, the BRST transformations are altered in a non-linear way beyond one loop. As we demonstrated in sec. \ref{sec:RGandCME} this brings the Legendre effective action and bare action realisations of the CME equation into tension with each other. This tension is resolved by the RG identities for a perturbatively non-renormalisable theory, which relate lower loop $\ell'<\ell$ counterterm diagrams to higher order poles at $\ell$-loop order \cite{Buchler:2003vw}. 

In this paper we only demonstrated how this works at two loops. A fully general understanding of how the RG ensures consistency for BRST seems possible, following the general understanding of the RG identities \cite{Buchler:2003vw} and \eg the proof of  renormalisability put forward in ref.  \cite{Gomis:1995jp} for effective theories with gauge invariance (the latter does not address the above tension but proceeds assuming both realisations of the CME remain consistent with each other).

Let us emphasise that the way the RG and BRST relations work together is quite remarkable. On the one hand the RG relates the two-loop double-pole vertices to the one-loop single pole vertices through a linear map which however involves computing further one-loop Feynman diagrams (the counterterm diagrams). On the other hand BRST, through the second-order CME relation \eqref{StwoCME}, directly relates the two-loop double-pole vertices to the square of the one-loop single-pole vertices, \ie without involving further loop calculations. In a sense then the BRST relations achieve what generalised beta function proposals, \cf sec. \ref{sec:solodukhin}, fail to do. 

However the BRST relations do not determine the higher pole vertices completely but only up to an $s_0$-closed piece, for example this is evident in the two-loop relation  \eqref{StwoCME}: $s_0S_{2/2}=-\frac12(S_{1/1},S_{1/1})$. They are thus less powerful than the RG identities. In fact in sec. \ref{sec:twoloops}, we saw in Feynman -- De Donder gauge that the non-linear term on the right hand side starts only at the three point level. As we explained, at the two-point level the equations degenerate, although they still allow a unique determination of the second order canonical transformations and thus also the new $s_0$-exact piece.

Let us note that the way the RG works to ensure consistency of BRST invariance, is not unique to non-renormalisable gauge theories. However in renormalisable theories, the divergent vertices are those in the original action. The RG identities for counterterm diagrams then play a less dramatic r\^ole in that they just ensure that these divergences appear with the correct sign so that they can be renormalised multiplicatively.

For quantum gravity, we verified the assertions above in sec. \ref{sec:Loops} by computing counterterms at one-loop up to the three-point level and up to two-loops for the graviton two-point vertex. Exploiting the BRST properties we gave a general parametrisation of the one-loop three-point counterterms and determined the parameters by matching to the graviton and ghost one-loop integrals. The antighost level two counterterms (which renormalise the BRST transformation of the ghosts) then follow without further Feynman diagram computations. 

These results could be readily extended, for example the ghost two-loop double-pole two-point counterterms can be computed using the vertices presented here and this would allow the form of the two-point $K_{2/2}$ to be fully determined, \cf eqn. \eqref{K2}. An interesting but more challenging project would be to work out the form of the one-loop counterterms to the next order in $\bar{h}_{\mu\nu}$ since this would allow one to determine the two-loop double-pole three-point background field vertices which would then allow a complete comparison with the off-shell results reported in ref. \cite{Goroff:1985th}. The parametrisation we give for $K_{1/1}$ in \eqref{K11} and \eqref{K21} looks sufficient to compute the corresponding one-loop counterterm diagrams, if the $d_i$ and $b_i$ terms are covariantised, however this introduces a number of new terms with undetermined coefficients, in particular we would need to determine the $h^*h^2\bR$ terms. The simplest way to do that would appear to be by matching to one-loop $h^*ch\bar{h}$ divergences. 

In our discussion of generalised beta functions in sec. \ref{sec:solodukhin}, we explained why they cannot be finite and verified this using our explicit results from sec. \ref{sec:Loops}. In particular for generalised beta functions based on the canonical transformations we obtained the formula \eqref{betadiverge} which shows why they cannot be finite. Nevertheless, this formula implies some interesting relations between the divergent higher order coefficients and the divergences generated by expectation values of the lower coefficients. It would be interesting to verify these and explore further their consequences. 

Finally let us return to our original motivation and note that the counterterms we have derived give directly the leading log behaviour at large euclidean  momentum. Indeed, the one-loop divergence \eqref{oneloopdiv} and counterterm \eqref{oneloop} taken together determine the $\ln(p^2/\mu^2)$ part. One can check explicitly that the two-loop double pole \eqref{twoloopdiags} from diagrams using only tree level vertices, together with divergences \eqref{counterloop} in one-loop counterterm diagrams and the double-pole counterterm from \eqref{twoloop}, conspire to cancel all but a remaining $[\ln(p^2/\mu^2)]^2$ term. Thus from the explicit results \eqref{oneloopzerotwopt} and \eqref{Stwotwo} we see that the leading log contribution of for example the two-point $h_{\mu\nu}$ vertex is given to two loops, in Feynman -- De Donder gauge, as:
\besp 
h_{\mu\nu}\Gamma^{\mu\nu\alpha\beta}(p)h_{\alpha\beta} = p^2\left(\ph^2-\frac12 h_{\mu\nu}^2\right)
+\frac{\kappa^2}{(4\pi)^2}\ln\!\left(\frac{p^2}{\mu^2}\right)\left(
\frac{61}{60}(R^{(1)}_{\mu\nu})^2-\frac{19}{120}(R^{(1)})^2 \right) \\
-\frac{\kappa^4p^2}{(4\pi)^4}\left[\ln\!\left(\frac{p^2}{\mu^2}\right)\right]^2\left(\frac{469}{7200} (R^{(1)})^2-\frac{79}{400}(R^{(1)}_{\mu\nu})^2+\frac{31}{1440}p^2 R^{(1)}\varphi\right)\,,
\eesp
where $h_{\mu\nu}$ and $\ph=\half h^\mu{}_\mu$ here just provide the polarisations, and the linearised curvatures \eqref{curvatures} should be similarly understood and cast in momentum space, thus $R^{(1)}_{\mu\alpha\nu\beta} = 2 p_{[\mu|\,}p_{[\nu} h_{\beta]\,|\alpha]}$ \etc 

Of course as physical amplitudes these corrections vanish on shell, while for the moment it remains just a dream that a way can be found to resum these leading contributions to all orders, where one might get powerful insights into the non-perturbative UV behaviour of quantum gravity. Nevertheless we hope that the detailed understanding we have gained of some of the consequences of combining background diffeomorphism invariance, RG invariance, and BRST invariance, bring that dream a step closer to reality.

\section*{Acknowledgements}
VMM and DS acknowledge support via STFC PhD studentships. TRM acknowledges support from STFC through Consolidated Grant ST/T000775/1.

\newpage
\appendix

\section{Comparisons with the literature}
\label{app:comparisons}

Here we outline the differences in convention and notation that need to be taken into account in order to compare with other results in the literature.

The two-point purely quantum one-loop counterterm given in \eqref{oneloopzerotwopt}, corresponding to the first diagram in fig. \ref{fig:twopoints}, was computed in a general two parameter gauge $(\tilde{\alpha}\partial^\mu h_{\mu\nu}+\tilde{\beta} \partial_\nu h^\rho_{\ \rho})^2$ in ref. \cite{Capper:1979ej}. (We put a tilde over his parameters so as not to confuse with the ones in this paper.) After taking into account the Minkowski signature and that factors of $1/(2\pi)^4$ are accounted for differently, it should coincide with the first two terms in \eqref{oneloopzerotwopt} on specialising $\tilde{\alpha}=1$ and $\tilde{\beta}=-\half$ to get Feynman DeDonder gauge. 

Initially the results did not coincide. Recomputing the two-point vertex in this general gauge we found the following typos in ref. \cite{Capper:1979ej}: in the square brackets of his $T_3$ there should be an extra term: $+\frac{45}8\tilde{\beta}^4/\tilde{\alpha}^2$, and in $T_4$ the term $-135(\tilde{\beta}^2/\tilde{\alpha})$ should read $-135(\tilde{\beta}^2/\tilde{\alpha}^2)$. Finally his parameter $a$ should be defined as $a=\half T_2-T_3$, rather than $\half E_4$ as stated. Once these are fixed, we find complete agreement.

The result for the purely quantum pieces in \eqref{oneloopzerotwopt} also agrees with the result quoted in ref. \cite{Kellett:2020mle} on recognising that there the divergence can be recovered by setting $\ln(1/\mu_R)=1/2\eps$. This mapping is also the one to use to compare the level one divergence with \eqref{onelooponetwopt}.

The purely background terms in \eqref{oneloopzerotwopt} agree with ref. \cite{tHooft:1974toh} on recognising that their $\eps=8\pi^2(d-4)$, their definition of Ricci curvature is minus ours, \cf below \eqref{dimreg}, and that their action is defined to be the opposite sign from the usually defined Euclidean action, \cf \eqref{minimal classical action}. Their normalisation of the scalar curvature term is also non-standard but this is repaired by mapping $g_{\mu\nu}\mapsto \sqrt{2}\kappa g_{\mu\nu}$ and has no effect on the one-loop result, since it is a curvature-squared action. 

In the famous paper \cite{Goroff:1985th}, this result is reproduced but the value quoted is half that of \eqref{oneloopzerotwopt}. To see this one should note that it is Minkowski signature and their $\eps=4-d$ \ie is twice ours. (There is also an accidental extra factor of $1/\eps$ in their quoted equation.) They also quote a value for some two-loop double-pole divergences. The one point of comparison is the result \eqref{Stwotwo} for the $\Bar{R}^{\mu\nu}\square \Bar{R}_{\mu\nu}$ counterterm. Using these translations we see that their result is again half of what we find.


\bibliographystyle{hunsrt}
\bibliography{references}


\end{document}